# In our mind's eye: Visible and invisible in quantum theory, with Schrödinger's cat experiment

Arkady Plotnitsky*

*Literature, Theory, Cultural Studies Program; Philosophy and Literature Program, Purdue University, West Lafayette, IN 47907, USA; Email: plotnits@purdue.edu

**Abstract**. This article aims to reconsider E. Schrödinger's famous thought experiment, the cat-paradox experiment, and its place in quantum foundations from a new perspective, grounded in the type of interpretation of quantum phenomena and quantum mechanics, which belongs to the class of interpretations designated here as "reality without realism" (RWR) interpretations. Such interpretations have not been previously brought to bear on the cat experiment, including by N. Bohr, whose interpretation in its ultimate form (as he changed his interpretation a few times) is an RWR interpretation, but who does not appear to have commented on the cat experiment. The interpretation adopted in this article follows Bohr's interpretation, as based on two assumptions or postulates, the Heisenberg and Bohr postulates, but it adds a third postulate, the Dirac postulate. The article also introduces, in conjunction with the concept of reality without realism, the concepts of visible and invisible to thought and considers their role in the cat-paradox experiment.

> Hamlet:
> My father — methinks I see my father.
> Horatio:
> Where, my lord?
> Hamlet:
> In my mind's eye, Horatio.
> --William Shakespeare, *Hamlet*, Act 1, Scene 2, ll. 183-185
>
> To die for the invisible. This is metaphysics.
> --Emmanuel Levinas, *Totality and infinity: An essay on exteriority*

## 1. Introduction

This article aims to reconsider E. Schrödinger's famous thought experiment, the cat-paradox experiment (hereafter "the cat experiment"), and its place in quantum foundations from a new perspective, which removes any paradox from it [Schrödinger 1935]. So, admittedly, do some other views of the experiment, and Schrödinger himself did not call it a paradox, rather a "ridiculous situation" [Schrödinger 1935, p. 157]. The present view of it, however, is grounded the type of interpretation of quantum phenomena and quantum mechanics (QM) that has not, to my knowledge, been previously brought to bear on the cat experiment, including by N. Bohr, whose interpretation, especially in its ultimate form, is the closest to the one adopted here, but who does not appear to have commented on the cat experiment. It is worth keeping in mind that, while an interpretation of QM, commonly, including Bohr's or the present interpretation, involves an interpretation of quantum phenomena, the latter have separate interpretable aspects (noted whenever necessary in this article) that do not depend on, and hence could be interpreted independently of, any theory predicting them. Quantum phenomena will be assumed here to be defined by the fact that in considering them (technically, the data found in them, pertinent to quantum experiment), the Planck constant, $h$, which is a classically measurable quantity, must be taken into account.[1]

---

[1] I put aside qualifications of this definition, necessary in general but not germane for this article, because all quantum phenomena and measurements considered involve $h$ (e.g., [Plotnitsky 2021a, pp. 37-38], also [Khrennikov 2021]). I might only add that all quantum-mechanical equations used for actually predicting the data observed in

Bohr eventually came to see quantum phenomena as revealing "a novel feature of atomicity in the laws of nature," "disclosed" by "Planck's discovery of the quantum of action [$h$], supplementing in such unexpected manner the old [Democritean] doctrine of the limited divisibility of matter" [Bohr 1938, p. 94]. Atomicity and, thus, discreteness or discontinuity initially emerged on this Democritean model, with M. Planck's discovery of the quantum nature of radiation in 1900, which led Planck to his concept of the quantum of action, $h$, physically defining this discontinuity, and then A. Einstein's introduction of the concept of a photon, as a particle of light, in 1906. The situation, however, gradually, especially with the discovery of QM in 1925 by W. Heisenberg, revealed itself to be more complex, eventually leading Bohr to his concepts of phenomenon and atomicity (essentially equivalent to that of phenomenon, but highlighting some of the features of the latter concept, such as discreteness), and the interpretation of quantum phenomena and QM based in these concepts.

This interpretation, developed in the later 1930s, became the ultimate version of Bohr's interpretation, following a decade of the development of, and some significant changes in, his views (with only a few minor refinements added later). This requires one to specify to which version of his interpretation one refers, which I shall do as necessary, while focusing on his ultimate interpretation, unavoidably, in the present interpretation of his interpretation. Unless qualified, "Bohr's interpretation" will refer to his ultimate interpretation. (The designation "the Copenhagen interpretation" requires even more qualifications as concerns whose interpretation it is, say, that of Heisenberg, Dirac, or von Neumann. Accordingly, I avoid this designation altogether.) The interpretation adopted in this article follows this interpretation, in particular as based on two assumptions or postulates, the Heisenberg and Bohr postulates, but it adds a third postulate, the Dirac postulate. All three postulates are defined below. I would like, however, to emphasize from the outset that these postulates are interpretive assumptions that could, in principle, be falsified, even though, as discussed later in this article, a falsification of an interpretation is not the same (and a more complex matter than) that of a theory by experimental evidence. By virtue of the first two postulates, especially the Heisenberg postulate, both interpretations belong to the class of interpretations of quantum phenomena and QM, or quantum field theory (QFT), designated here as "reality without realism" (RWR) interpretations. This article is only concerned with QM and, marginally, QFT (in high-energy regimes) in their currently standard forms, and puts aside, except in passing, alternative quantum theories, such as Bohmian mechanics or spontaneous collapse theories.[2]

The Heisenberg postulate, most essentially defining RWR interpretations, was in effect introduced by Bohr's 1913 atomic theory, in considering the transitions, "quantum jumps," between stationary states of electrons, while retaining a realist view of stationary states by assuming them to be represented as orbits of electrons around nuclei. The RWR understanding of quantum phenomena and QM emerged in its full form in Heisenberg's approach to quantum theory that led him to his discovery of QM, which is why I use the designation "the Heisenberg postulate." The Heisenberg postulate places the emergence of quantum phenomena beyond representation or knowledge, or even conceptions, beyond the reach of

---

quantum phenomena contain $h$ or $\hbar$ (or something mathematically equivalent, for example, by suitable changing the values of the parameters, such as time), which fact is sometimes hidden in more abstract, such as Hilbert-space, versions of the formalism, unless one properly unfolds its relevant elements to make actual predictions possible.

[2] The interpretation offered in this article was considered previously in [Plotnitsky 2021a,b; 2022a,b]. The last article cited expressly adopts the three postulates in question, under the headings of Heisenberg, Bohr, and Dirac *discontinuity*, in considering the double-slit experiment. RWR interpretations without the Bohr postulate may be possible, but they will be put aside here, because both Bohr and the present interpretation adopt this postulate, along with the Heisenberg postulate, which defines all RWR interpretations. It is possible and technically more rigorous to see a different interpretation of a given theory as forming a different theory, because each interpretation may involve concepts not be shared by other interpretations. This is the case, for example, in different versions of "the Copenhagen interpretation," not all of which are RWR interpretations, which too may be different, as are Bohr's and the present interpretation, because the present interpretation assumes the Dirac postulate. What is shared is the mathematical formalism used, at least in terms of the equivalence or mutual translatability of its different versions. For simplicity, however, I shall continue to speak of different interpretations of a theory itself containing a given mathematical formalism, specifically, of different interpretations of QM or QFT.

thought, or in terms I shall adopt here, make this emergence *invisible to thought*. Realism, by contrast, is defined by the assumption of the possibility of either representation or knowledge, or at least conception of how the phenomena considered are possible, thus making them *visible to thought*. I shall speak of *weak* RWR interpretations (or the weak form of the Heisenberg postulate) when this emergence is assumed to be beyond representation or knowledge, and of *strong* RWR interpretation (or of the strong form of Heisenberg postulate) when it is assumed to be beyond conception and thus made invisible to thought. This article adopts a strong RWR interpretation, as did Bohr in the ultimate version of his interpretation. Unless qualified, the term "RWR interpretation" will, hereafter, refer to strong RWR interpretations.

The concepts of classical physics, specifically classical mechanics, emerged as mathematized refinements of our daily concepts—concepts arising from our general phenomenal experience of the world. All modern physics (classical, relativistic, or quantum) only deals with suitably mathematized idealizations of physical reality. This connection between physical and daily concepts has proven to be difficult to use in quantum theory, even in realist interpretations, which would assume that QM or QFT provides a mathematized representation of quantum objects and processes. Such a representation is no longer a mathematical refinement of our general phenomenal experience of the world, although QM or QFT formally adopts some (but only some) mathematics used in classical physics. RWR interpretations preclude any representation, including any mathematical representation, or even conception of the *ultimate* reality responsible for quantum phenomena. I qualify because classical physics remains an essential part of quantum theory, including in RWR interpretations, if they assume, as both Bohr's and the present interpretation do, the Bohr postulate. By the Bohr postulate, quantum phenomena, defined by what is observed in measuring instruments, along with the observable parts of these instruments, are represented by classical physics.[3] By classical physics I mean (as Bohr appears to have done, although he did not always specify the term) to classical mechanics and classical electromagnetic theory, with the addition of special relativity in high-energy (QFT) quantum regimes. By classical mechanics I refer to Newton's mechanics defined by its three main laws, the law of inertia, the law of the changes a force can have on the motion of a body, $F = ma$, and the law of action and reaction between interacting bodies, as equal in magnitude by opposite in direction. These three theories—classical physics, *relativity*, and quantum theory—are sufficient for the present purposes for representing the observable part of measuring instruments in quantum physics. (Gravity, governed by Newton's forth law, is a special case of Newton's mechanics, which can be put aside for the present purposes, as can be the equivalence principle, not involved in any phenomena considered in this article.) Accordingly, by a quantum instrument I understand any technological device able of establishing quantum phenomena and registering quantum data, the data that is represented by classical physics but that cannot be predicted by classical physics (including as concerns the role of $h$ in these data), and thus requires an alternative, quantum theory, such as QM or QFT, of possibly some alternative theory. (This article, again, will put such alternatives, be they actual or hypothetical, aside.) At least, such is the case, *as things stand now* (a qualification assumed throughout this article), for it is in principle possible that one or another form of classical theory able to predict this data can be developed, and there are attempt in this direction, which will be put aside here, because this article is only concerned with RWR interpretations of QM or QFT and, a subject not considered previously, how the cat paradox appears in this interpretation. I make no other claims here. Measuring instruments, it follows, also have quantum strata through which they interact with quantum objects. Eventually, Bohr adopted the term "phenomenon" to refer strictly to what is observed in measuring instruments, as effects of their interaction with quantum objects [Bohr 1987, v. 2, p. 64]. The Bohr postulate, thus, also reflects the transition, via measuring instruments, from the ultimate, "quantum," reality considered to the classical reality of observation, and conversely, in the initial stage (preparation) of an experiment, from the classical reality of observation to the ultimate, "quantum," reality considered. In strong RWR interpretations, when referring to this ultimate reality, "quantum," or, in the first place

---

[3] RWR interpretations on the Heisenberg postulate alone (defining all RWR interpretations), without assuming the Bohr postulate, are possible. Conversely, the Bohr postulate need not be limited to RWR interpretations and is found in realist interpretations of quantum phenomena and QM.

"reality," is a term to which no concept we can form can be associated. Any such association is only possible and necessary at the level of observation or phenomena. At this level, that of visible to thought or even available to immediate human perception, thus defining quantum phenomena or quantum events, the term "quantum" has (classical) physical concepts, such as discreteness or individuality, associated with it.

By the Heisenberg postulate, how quantum phenomena come about cannot, in RWR interpretations, be represented by QM or QFT, but only predicted by it, in general probabilistically. QM or QFT, has, in these interpretations, no physical connections apart from making these predictions, to either the ultimate nature of reality responsible for quantum phenomena or, because they are described by classical physics, to these phenomena themselves. Hence, in these interpretations, the capacity of the mathematics of QM or QFT to predict the outcomes of quantum experiments, even if only probabilistically (which is, however, in accord with the experimental evidence now in place), becomes in turn beyond knowledge or even conception. We know *how* this mathematics works (how to use it), but we do not know and perhaps cannot know or even conceive of *why* it works. Fortunately for us, however, it does work.

By contrast, in classical physics or relativity (special or general), the mathematical formalism (ideally) represents, make visible to thought, the physical reality responsible for the phenomena considered and connects, by continuous processes, these phenomena. The latter can, moreover, be identified with the physical objects considered, because the interference of measuring instruments can be neglected for all practical purposes. This identification is no longer possible in considering quantum phenomena, in the constitution of which the role of measuring instruments is irreducible. Nobody has ever *seen* a moving electron or photon. It is invisible to an observation and, in the RWR view, even *invisible to thought*, is entirely beyond the reach of thought. It is only possible to observe traces, which are visible even to our immediate sense perception and consciousness (such traces may also be "clicks" that we hear rather than see) of their interactions with measuring instruments. These traces make it difficult and, in strong RWR interpretations, impossible to reconstitute the ultimate nature of the reality responsible for them, whether one *sees* this reality in terms of quantum objects or assumes, as I do here, that a quantum object is an idealization applicable only at the time of measurement. (This assumption is the content of the Dirac postulate.) Either way, this situation entails an unavoidable discrimination between quantum objects and instruments, and hence phenomena, a discrimination that, according to Bohr, "may indeed be said to form a *principal distinction between classical and quantum-mechanical description of physical phenomena*" [Bohr 1935, p. 701].[4]

While adopting this structure of observation in quantum physics, the present interpretation further stratifies it by the Dirac postulate, not found in Bohr. Bohr's argumentation might be seen as, at certain points, suggesting the Dirac postulate. Bohr, however, never formulated this type of postulate or the corresponding view and appears to have always assumed that the concept of a quantum object is an entity that exist independently, while still being beyond representation or even conception, by the Heisenberg postulate. According to the Dirac postulate, the concept of a quantum object is only applicable at the time of observation, but not to anything assumed to exist independently in nature. In Bohr's interpretation in all its versions, the ultimate reality responsible for quantum phenomena was associated with quantum objects, eventually as independent RWR-type entities, different from quantum phenomena, defined by the irreducible role of measuring instruments, the observable parts of which are described by classical physics. A quantum object is, in Bohr view, a physical object responsible for the existence of a quantum phenomenon, as an effect of the interaction between this object and a measuring instrument or some (classical) object existing in nature that function as an instrument for us. Nothing, either built by us or by nature, can be defined as an instrument apart from us in the present and, I would argue, in Bohr's view. As RWR type entities quantum objects were assumed by Bohr to be beyond conception, and hence could not be assigned any properties, including *h*, even at the time of observation and measurement, as all physical properties were only assignable to observables parts of measuring instruments, described classically. As will be seen, it is possible in a quantum experiment to consider as *the object under investigation* an object that also contains a classical part, such as the cat in the cat experiment, but this composite object must still contain a properly quantum object for the observed phenomenon to be a quantum phenomenon. QM would predict such observed properties of measuring instruments and only them, rather than any properties of quantum objects. The Dirac postulate introduces a triple rather the double, stratification into this situation, following [Plotnitsky 2021a, 2022a,b]. The ultimate RWR reality responsible for quantum phenomena is an idealization assumed to exist independently of our interactions with it, and thus independently of observation. By contrast, the concept of a quantum object, elementary, such as a photon or electron, or composite (possibly macroscopic) is an idealization that, while still of the RWR-type, only applies *at the time of an observation*. An observation becomes a creation of a quantum phenomenon by the interaction between the ultimate RWR-type reality, and the instrument we use, and the capacity of our thought to observe the phenomena thus created, which also allows enables one to apply the concept of quantum object, by the Dirac postulate only after an observation has taken place. In all three cases—the (independent) ultimate RWR reality responsible for quantum phenomena, quantum objects, and quantum phenomena—one only deals with idealizations created by our thought.

---

[4] The difference between phenomena and objects has its genealogy, in modern times (it had earlier precursors, even in ancient Greek philosophy), in I. Kant's distinction between objects as things-in-themselves in their independent existence and phenomena as representations created by our mind, which may not correspond to the objects which they are aiming to represent or to which they may be representationally unrelated at all [Kant 1997]. The latter is in fact the case in considering quantum phenomena vis-à-vis quantum objects because quantum phenomena represent classical physical objects observed in measuring instruments. As the strong RWR view, Bohr's or the present view is more radical than that of Kant. While Kant's things-in-themselves are assumed to be beyond knowledge, they are not beyond conception, at least a *hypothetical* conception, even if such a conception cannot be guaranteed to be correct and is only practically justified in its applications [Kant 1997, p. 115]. By contrast, in the strong RWR view what is practically justified is not a possible conception of the ultimate nature of reality responsible for quantum phenomena, but the impossibility of such a conception, thus in precluding this reality from being visible to thought, even hypothetically. No other justification than practical is possible by virtue of the impossibility of this conception. The concept of a quantum object can of course be considered from alternative, including realist, perspectives. See, for example, [Jaeger 2014], which offers a rigorous argument for such a concept.

The reason for using the designation "Dirac postulate" is that, while, unlike the Heisenberg postulate by Heisenberg and the Bohr postulate by Bohr (even if without using these designations as such), this postulate was not considered by Dirac himself, it may be seen as having emerged from Dirac's famous equation for a relativistic electron. While originally written for an electron, Dirac's equation

$$\left(\beta mc^2 + \sum_{k=1}^{3} \alpha_k p_k c\right)\psi(x,t) = i\hbar \frac{\partial \psi(x,t)}{\partial t}$$

$$\alpha_i^2 = \beta^2 = I_4$$

($I_4$ is the identity matrix)

$$\alpha_i \beta + \beta \alpha_i = 0$$
$$\alpha_i \alpha_j + \alpha_j \alpha_i = 0$$

revealed itself to an equation for both the (free) electron and the (free) positron, including their spins, which the equation contains automatically, in contrast to QM, where predicting the spin of an electron needs to be handled separately, via Pauli matrices combined with Schrödinger's equation. Dirac's equation reflected and, as it happened, led to the discovery that a different particle (in the present view, defined in terms of effects observed in measuring instruments) can be registered in a single experiment: the initial observation can register an electron, while the next one a positron, or a photon, or an electron-positron pair, with the probabilities defined by the same equation. Once one moves to still higher energies, the panoply of possible outcomes becomes even greater. In QED, one only deals with electrons, positrons, and photons; in QFT, depending how high the energy is, one can find any known elementary particle or combination, that is, the corresponding effects will be registered. Accordingly, it is reasonable to apply the concept of a quantum object (still as an RWR-type entity) exclusively at the time of observation. There are, however, reasons to adopt this view in low-energy (QM) quantum regimes, including in order more effectively to interpret quantum conundrums, such as that of the double-slit experiment, a paradigmatic and, arguably, the most famous quantum experiment [Plotnitsky 2022b].[5]

Do quantum phenomena or QM *require* the Dirac postulate, or the Heisenberg and Bohr postulates? It would be difficult to argue such a case, and it is not my aim to do so. My only claim is the logical consistency of the interpretations, such as the one adopted here, grounded in these postulates, and their accord with the experimental evidence currently available. As I said, new experimental evidence can change the present situation of fundamental physics, just as then new evidence changed its situation around 1900, leading to Planck's discovery of quantum theory.

The next section outlines the (RWR-type) interpretation adopted in this article, cast in terms of the relationships between visible and invisible to thought. Section 3 discusses, by way of the bridge to the cat experiment, the letter exchange between Schrödinger and Bohr (at the time Schrödinger's work on his paper containing the cat paradox) concerning the use of classical concepts in quantum measurement, thus, essentially the Bohr postulate. Section 4 considers Schrödinger's cat experiment from the (RWR) perspective established by the preceding analysis. The conclusion offers philosophical reflections on the role of metaphysics in physics, via the relationships between visible and invisible to thought.

## 2. Reality without realism, and visible and invisible to thought in fundamental physics

This section outlines the RWR *view* of quantum phenomena and quantum theory, a view cast in terms of the relationships between visible and invisible to thought. For simplicity, I shall primarily discuss QM, only briefly referring to QFT, although my argument applies to and can be further supported by QFT. I speak, more generally, of "the RWR view" because it can lead to various interpretations. These interpretations share the Heisenberg postulate, defining RWR interpretations, but beyond being either weak or strong RWR interpretations, some of them may contain additional postulates, such as the Bohr postulate or the Dirac postulate. Thus, while Bohr's interpretation and the present interpretation both assume the Bohr postulate, only the present interpretation assumes the Dirac postulate.

---

[5] One can translate the argument of this article into quantum-informational terms. In the present view, information is human. Nature has no information, only we do, possibly about nature, or what we assume to be nature. All information obtainable in quantum experiments is contained in the data observed in measuring instruments, described classically, by the Bohr postulate. Hence, this information qua information is classical, Shannon information (measured in classical bits), and as such is visible to thought and communicable unambiguously given the mathematical nature of Shannon information (as opposed to other forms of information with a semantic content, which may allow for ambiguity). However, this information and its organization, as manifested in quantum experiments cannot be predicted or processed by classical means. The emergence of this information requires the assumption of quantum objects or in the present view, by the Dirac postulate, an RWR type reality ultimately responsible for quantum phenomena and the use measuring instruments capable on interacting with this reality, and a theory, such as QM, different from classical theories, that is capable of handling this information. In this sense, while all actual information obtained in experiments is classical, one can speak, as is common, of "quantum information." In the present view, the "quantum," as the ultimate reality responsible for quantum phenomena, in only a particular way, defined by out interaction with nature, to create and communicate classical information, which cannot be predicted by classical physics (or relativity) and the structure of which cannot be generate by classical objects and processes. Dealing with quantum information requires quantum information science, a vast subject of its own, including as concerning its impact on quantum foundations. Quantum information theory brings new features to the differences between classical and quantum information in relation to QM or QFT, because at this stage quantum information theory is primarily concerned with discrete variables, physically represented by spin, rather than continuous variables. While there are classical and quantum versions of continuous variables, spin is a quantum variable, corresponding to a strictly quantum aspect of nature, and has no classical analog. By the same token, a spin may be seen as reflecting something in nature strictly beyond our thought's capacity to form a conception of, a strictly RWR-type entity. I am indebted to G. M. D'Ariano for exchanges concerning of the visible and communicable or (his preferred term) "sharable" nature of classical information. I am not claiming that he subscribes to the argument of this article or all of this argument, which builds on [Plotnitsky 2021a,b, 2022a,b], as concerns the RWR view and the idea of the invisibility to thought, as extending to *all thought*, including mathematical and physical thought, rather than only our immediate (conscious) phenomenal intuition or visualization. In general, this article's argument is independent of quantum information theory.

The philosophical position grounding the present interpretation implies that modern physics, as a mathematical-experimental science, contains two forms of thinking, which may be designated as "classical" and "quantum" in view of their respective origins in classical and quantum theory. Both assume the physical reality they consider to exist independently of our existence and thinking as humans, but each treats this reality differently as concerns the *ultimate nature* of this reality, assumed to be representable and, thus, visible to thought in classical thinking, and to be beyond not only representation but also conception and, thus, invisible to thought in quantum thinking. I speak of the ultimate nature of the reality considered or (for the sake of economy) just the ultimate reality considered, because quantum thinking assumes that classical thinking applies at some levels of the reality considered, specifically, by the Bohr postulate, that of the observable parts of measuring instruments. Quantum thinking also involves classical thinking. In adopting only classical thinking one assumes it to apply at all levels of physical reality, without allowing for quantum thinking. These two forms of thinking are as follows:

(1) *Classical thinking*, which is essentially a realist thinking, deals with a form of reality that is *visible to thought*, as what can be perceived, imagined, visualized, represented, known, conceptualized, and so forth, and as such allows one to have statements or images of this reality that can, at least in principle, be communicable unambiguously, as is necessary for the practice of science, as constituted now;

(2) *Quantum thinking*, which is essentially RWR thinking, contains classical thinking but also assumes the existence of a form of reality, as the ultimate reality considered, that is no longer available to classical thinking and, as such, is *invisible to thought,* as what cannot be perceived, imagined, visualized, represented, known, conceptualized, and so forth, which also means that nothing about it can be communicated unambiguously, if at all, apart from the claim that it is beyond the reach thought.[6]

Thus, in the case of this (RWR) form of reality, quantum thinking, divorces the term "reality" from any possible concept associated to it, making it akin to a mathematical symbol, like *R* or *X,* which could have been used instead of the word reality here. "Reality without realism" or RWR functions in this way as well. I emphasize that quantum thinking also assumes forms of reality, such as that observed, as phenomena, in quantum experiments, that handled by classical thinking by the Bohr postulate and as such is visible to thought. The very existence of any RWR-type reality is inferred from certain configurations (such those defining the data observed in quantum experiments) of classical reality.

Classical and quantum *thinking* are not the same as physical *theories* or *interpretations* using either thinking, because such theories contain additional features. Thus, while different theories, both classical physics and relativity, special or general, conform to classical thinking, which quantum theory does not at least in RWR interpretations. It is true that special relativity severely limits the capacity of *our immediate phenomenal intuition* to represent or visualize the kinematic used by the theory. These qualifications, however, do not prevent this kinematic from being *visible to thought* and allow for a realist treatment, because relativity, special or general, represents all reality considered in it in terms of suitably mathematized physical concepts, just as does classical physics. In considering classical physical phenomena or (they can, again, be identified with each other in classical physics or relativity) objects the role of both *h* and *c* can be disregarded and is by classical mechanics; in considering relativistic phenomena or (they can, again, be identified with each other) objects, only *h* can be disregarded; and in considering quantum phenomena or (they can no longer be identified with each other) objects, *h* be taken into account, and in high-energy relativistic (QFT) regimes, $c$ must be taken into account as well.[7] This formulation is different from saying that $c$ is assumed to be infinite in classical physics and $h$ equal to zero in classical physics and relativity. In particular, classical mechanics need not be, and in the present view is not, assumed to be the limit of QM by putting $h$ equal to zero or express in this way Bohr's correspondence principle (explained below). These are, in the present view, two different theories, dealing with two different types of objects, even though both are ultimately composed of quantum objects or (in the present view) the same ultimate reality: classical mechanics is not a special (limit) form of QM and classical objects are not a special type of quantum objects, although quantum objects can sometimes be treated classically, which claim, however, requires important qualifications explained below. In the present interpretation, moreover, quantum objects are only defined at the time of observation by the Dirac postulate, which does not apply to classical objects, defined independently of observation.

---

[6] By thus relating the concepts of "visible to thought" and "unambiguously communicable," specifically, by means of language (although language is not the only form of unambiguous communication, which can be visual, for example), *speaking* of visible and invisible to thought suggests a connection to J. S. Bell's title *Speakable and Unspeakable in Quantum Mechanics* [Bell 2004], a collection of his writings, primarily on QM. The philosophical position adopted in this article is, however, opposite of that of Bell, which also leads Bell to his discontent with Bohr's view, including the Bohr postulate, and with QM in the first place, as discussed in [Plotnitsky 2021b, 2022b].

Quantum thinking emerged in quantum theory in RWR interpretations in view of the nature of quantum phenomena, assumed to be the effects on the interactions between the ultimate reality responsible for these phenomena and suitable measuring instruments. These phenomena, or rather numerical data they contain, are predicted by quantum theory, QM or QFT, without, in RWR interpretations, representing the ultimate reality responsible for them. QM or QFT does not represent these effects either. They are, by the Bohr postulate, represented by classical physics, which enables them to be as unambiguously communicable, as is the mathematics of QM or QFT. Along with the predictive capacities of both theories, the possibility of this unambiguous communication and, in this sense (a qualification discussed below), objectivity make these theories conform to "the basic principles of science," as stressed by Bohr (e.g., [Bohr 1935, p. 700; Bohr 1987, v. 2, pp. 67-68, v. 3, p. 7]). On the other hand, classical theories cannot predict these effects. It follows that both types of theories are necessary in fundamental physics, including quantum theory, because, by the Bohr postulate, quantum phenomena are represented by classical physics, with adding special relativity in high-energy (QFT) regimes, while they can only be predicted by quantum theory, which in RWR interpretations represents neither these phenomena not, by the Heisenberg postulate how they come about. In fact, classical physics is necessary for describing measuring instruments in relativity as well, specifically in all measurements within each local reference frame, even though the instrument are subjects to the relativistic laws of motion. While philosophically classical, that is, conforming to classical thinking in the sense defined here, general relativity is a separate part of fundamental physics as currently constituted. It may sometime play a role in dealing with quantum phenomena, keeping in mind that the emergence of quantum phenomena considered thus far does not involve gravity as a such, as we do not have a quantum theory of gravity.[8]

Classical theories are grounded, in Bohr's words, "the idea [and hence the assumption] that the phenomena concerned may be observed without disturbing them appreciably," which enables one to identify these phenomena with the objects considered [Bohr 1987, v. 1, p. 53]. This assumption no longer appears possible in considering quantum phenomena, empirically and hence regardless of interpretation. In fact, it may not be rigorously possible even in classical physics and relativity, insofar as all phenomena considered are still created by our thought, which is a product of our bodies and brains, as experimental technologies created by nature [Plotnitsky 2021a, pp. vii-xxiv]. However, the assumption that "the phenomena concerned may be observed without disturbing them appreciably" is workable in these theories for all practical purposes. By contrast, in Bohr's or the present interpretation, quantum theory, QM or QFT, are defined by dealing with the combination of fours features:

> (1) the ultimate, "quantum," reality responsible for quantum phenomena, a reality invisible to thought by the Heisenberg postulate and commonly, including by Bohr, identified with quantum objects, which are, however, by the Dirac postulate, defined (still as RWR-type entities and thus invisible to thought) only at the time of observation in the present interpretation;
> (2) observational technology, commonly understood as comprised of measuring instruments;
> (3) observed phenomena, created by the interaction between quantum objects and measuring instruments, phenomena that are always visible to thought and even to our immediate phenomenal perception, while the numerical data observed and, in the first place, the observable parts of measuring instruments, are described by classical physics by the Bohr postulate;
> (4) the mathematical formalism of QM (cum Born's rule), probabilistically or statistically predicting the outcomes of quantum experiments, as observed, via measuring instruments, in quantum phenomena, without representing the ultimate reality responsible for them by (1).

---

[7] See, however, note 1 above.

[8] It may in principle be contended that, even if necessary for the description of the observable phenomena in quantum or relativistic measurements, classical physics is not a separate theory, for example, by assuming or arguing that ultimately all physical objects considered are quantum. In the present view, however, classical physics is a necessary separate part of fundamental physics. It may not, as such, deal with the ultimate constitution of matter, but I would argue, it is unavoidable in considering many, even most, macroscopic phenomena, and, again, even in dealing with fundamental, such as quantum, physics, where classical physics is unavoidable in dealing with observation and measurement. Given that gravity plays no role quantum (or classical) phenomena considered in this article, I put it aside, although it may ultimately bear on the set of questions posed here.

Measuring instruments, it follows, contain both classical, observable, strata of reality and unobservable, ultimately invisible-to-thought, quantum strata of reality, which enables this interaction.

The reasons for my emphasis on visible and invisible, extended to the idea of visible and invisible to thought (essentially thinkable and unthinkable) are both conceptual and historical. Conceptually, our capacity for visualizing the world, which is also related to the neurological functioning of our brain (about 60% of which is dealing with vision), is crucial to our thought. This capacity has shaped classical physics, as a mathematical refinement of the world we observe, but it was defeated, first by special relativity, the kinematic of which is beyond it, and then, more radically, by quantum theory, which brought into physics that which is invisible to thought altogether, is beyond thought.

Historically, this emphasis follows Bohr's appeal to the impossibility of visualization of the ultimate responsible for quantum phenomena, defined as effects of the interaction between this reality and our agencies of observation. Bohr's use of visualization and its avatars was in part shaped by the German term for intuition, *Anschaulichkeit*, which etymologically relates to what is phenomenally visualizable. Even before (albeit only by a few months) Heisenberg's discovery of QM, based on "abandoning the ordinary spacetime description," and hence on an RWR type view, at least in its weak form [Bohr 1987, v. 1, p. 48], Bohr said:

> I am forcing myself these days with all my strength to familiarize myself with the mysticism of nature and am attempting to prepare myself for all eventualities, indeed even for the assumption of a coupling of quantum processes in separated atoms. However, the cost of this assumption are so great that they cannot be estimated within the ordinary spacetime description. [A Letter to Heisenberg, April 18, 1925, Bohr 1972–1996, v. 5, pp. 79–80, p. 237]

In a letter to Born, a few days later, he added:

> [Quantum experiments] preclude the possibility of a simple description of the physical occurrences [at the quantum level] by means of visualizable pictures . . . [S]uch pictures are of even more limited applicability than is ordinarily supposed. This is of course almost a purely negative assertion, but I feel that . . . one must have recourse to symbolic analogies to an even greater extent than hitherto. Just recently I have been racking my brain to dream up such analogies. [Letter to Born, 1 May 1925, Bohr 1972–1996, v. 5, p. 311]

The word "mysticism" will soon disappear from Bohr's writings, replaced by an emphasis on QM as a rational theory, free from any "mysticism incompatible with the true spirit of science" [Bohr 1937, p. 83, Bohr 1987, v. 2, p. 63]. By referring to "the assumption of a coupling of quantum processes in separated atoms," the statement also captures the core of the dilemma later posed by the Einstein-Podolsky-Rosen (EPR) experiment [Einstein et al 1935]. Bohr links this dilemma to the impossibility of visualization, ultimately making how what is observed there, or in any quantum experiment, come about invisible to thought. What makes Bohr's statements remarkable is that they were made in 1925, 10 years before EPR's article. It is true that Einstein brought up related considerations in his exchanges with Bohr already in 1927, which was, however, still two years away in 1925, and followed the invention of QM [Bohr 1987, v. 2, pp. 41-58].

There are numerous invocations of the limits and in effect the impossibility of visualization throughout Bohr's writing, with an increasing emphasis, ultimately amounting to dealing with invisible to thought, even if without using this language.[9] To cite some key passages, proceeding chronologically:

* "In atomic problems we have apparently met with such a limitation of our usual means of *visualization*" (1925) [Bohr 1987, v. 1, p. 51];

* "On the whole, it would scarcely seem justifiable, in the case of the interaction problem, to demand a *visualization by means of ordinary space-time pictures*. In fact, all our knowledge concerning the internal properties of atoms is derived from experiments on their radiation or collision reactions, such that the interpretation of experimental facts ultimately depends on the abstractions of radiation in free space, and free material particles. Hence, our whole space–time view of physical phenomena, as well as the definition of energy and momentum, depends ultimately upon these abstractions" (1927) [Bohr 1987, v. 1, p. 77];

* "The resignation with regard to the desires for *visualization* which gives our whole language its character, to which we are compelled by the situation [in QM]" (1929) [Bohr 1987, v. 1, p. 98];

* "Indeed, only a conscious resignation of our usual demands of *visualization* and [classical] causality] was it possible to make Planck's discovery fruitful in explaining the properties if the [chemical] elements of the basis of our knowledge of the building stones of atoms" (1929) [Bohr 1987, v. 1, p. 108];

* "We must only be prepared for the necessity for the necessity of ever extending abstraction from our customary demands for a *directly visualizable description of nature*. Above all, we might expect new surprises in the domain [of QED and QFT] where the quantum theory meets with the theory of relativity and where unsolved difficulties still stand as hindrance to a complete fusion of the extension of our knowledge and the expedients to account for natural phenomena which these theories have given us" (1929) [Bohr 1987, v. 1, p. 108];

* The fundamental indeterminacy which we meet here [in Heisenberg's uncertainty relations] may, as the writer [Bohr] has shown, be considered as a direct expression of the absolute limitation of the applicability of *visualizable conception* in the description of [the ultimate reality of] atomic phenomena. … The resignation as regards *visualization* and [classical] causality, to which we are thus forced in our description of atomic phenomena, might well be regarded as a frustration of the hopes which formed the [Democritean] starting-point of atomic conceptions. Nevertheless, from the present standpoint of the atomic theory, we must consider this very renunciation as an essential advance in our understanding" (1929) [Bohr 1987, v. 1, pp. 114-115];

* "Only [the] limitation of our *visualizable conception* of motion, which is characteristic of quantum theory, enables us to understand how electrons can make their way between the metal atoms in the wire" (1929) [Bohr 1987, v. 1, pp. 118];

* "We must be prepared for a more comprehensive generalization of the complementary mode of description [in QFT] which will demand *a still more radical renunciation of the usual claims of so-called visualization*" (1937) [Bohr 1937, p. 88];

* "The extent to which *renunciation of the visualization* of atomic phenomena [technically, how they come about] is imposed upon us is strikingly illustrated by the following example to which Einstein very early called attention and often has reverted [in effect that of the alternative, complementarity, behavior of photon, depending up the two alternative set-up of the experiment, in essence equivalently to the double slit experiment" (1949) [Bohr 1987, v. 2, p. 51];

* "… [T]he ingenious formalism of quantum mechanics … *abandons pictorial representation* and aims directly at the statistical account of quantum processes" (1951) [Bohr 1998, p. 152];

* "Indeed, *renouncing pictorial description of electronic constitution* of the atomic system and only making use of empirical knowledge of threshold and binding energies of molecular processes, we can within a wide field of experiences treat the reaction of such systems by ordinary chemical kinetics, based on the well-established laws of thermodynamics" (1962) [Bohr 1987, v. 3, p. 25].

---

[9] I have considered this aspect of Bohr's argumentation in [Plotnitsky 2012, 2016, 2021a].

* "Certainly the issue [raised by the EPR experiment] is of a very subtle character and suited to emphasize how far, in quantum theory, we are *beyond the reach of pictorial visualization*" [Bohr 1987, v. 2, p. 59; emphasis added].

The statement closing my traversal is Bohr's 1949 comment on the EPR experiment. As already Bohr's 1925 statements cited above make clear, for Bohr the cost of "the assumption of a coupling of the processes in separated atoms," at stake in the EPR experiment (which deals with two spatially separated quantum objects) was the impossibility of "the ordinary spacetime description" of how phenomena is observed there and, as such, described in space and time. The EPR experiment and, in fact, all quantum experiments "preclude the possibility of a simple description of the physical occurrences [of phenomena considered] *by means of visualizable pictures*" [Letter to Born, 1 May 1925, Bohr 1972–1996, v. 5, p. 311]. Bohr, thus, assumed that such may be the case well before EPR's article, which might be one of the reasons why he thought that EPR's experiment didn't contain anything essentially new. He might not have been entirely right on this point, given the role of entanglement and correlations brought about by the EPR experiment. EPR, however, did not consider these concepts either. That of entanglement was introduced by Schrödinger in response to EPR's paper, including in the cat-paradox paper [Schrödinger 1935, p. 161]. Correlations became prominent even later. Be it as it may on that score, Bohr argued that, although "the issue [raised by the EPR experiment] is of a very subtle character and suited to emphasize how far, in quantum theory, we are *beyond the reach of pictorial visualization*," EPR's argument does not demonstrate, as EPR claimed, either the incompleteness of QM or else its nonlocal nature (in the sense of allowing an instantaneous action at a distance). My main point at the moment is Bohr's view that the ultimate reality responsible for quantum phenomena is invisible to thought, a reality defined here as a reality without realism (RWR), while quantum phenomena, as, by the Bohr postulate, observed classically, are visible to thought or in fact available to our immediate phenomenal perception.

It might be useful to briefly consider, as a simple representative example, which illustrate and will help to guide my discussion of RWR interpretations, how predicting the polarization of a photon appears in these interpretations. There are two possible outcomes of measurement (after the initial preparation): for example, the horizontal state $x$ and the vertical state $y$, observed classically by the Bohr postulate. In RWR interpretations, one could not say, as it is said sometimes, that before it is measured, the photon is (or is prepared) in a superposition of two physical states, and in the present view, moreover, the very concept of a photon, while it cannot be observed as such (only the corresponding effect in measuring instruments can) is only applicable at the time of observation by the Dirac postulate. The wave function allowing one to predict either physical state $x$ or $y$ is written as $|\psi\rangle = \alpha|X\rangle + \beta|Y\rangle$ with probability amplitudes of $|\psi\rangle$ associated with state vector $|X\rangle$ given by $\alpha$ and $|Y\rangle$ given by $\beta$. In a random experiment, the probability of the photon, when its polarization will be measured, to be horizontally polarized is $|\alpha|^2$ and to be vertically polarized is $|\beta|^2$ (by Born's rule). (Actual predictions will involve $h$, which does not appear in these abstract notations, but will once there are properly unfolded to make actual predictions possible.) That, however, need not, and in the RWR view does not, mean that $|\psi\rangle = \alpha|X\rangle + \beta|Y\rangle$ represents the photon in a superposition of two *physical* states, $x$ and $y$**,** as nothing can be said, by the Heisenberg postulate, concerning what happens between observations in the RWR view. Only the mathematical state vectors, designated $|X\rangle$ and $|Y\rangle$ (in capital letters), in the Hilbert space used, are in a linear (mathematical) superposition, with given amplitudes, and not quantum objects, let alone the outcomes of experiments.

QM, then, in Bohr's or the present interpretation, does not represent either, by the Heisenberg postulate, the physical emergence of quantum phenomena or, by the Bohr postulate, the observed quantum phenomena, represented by classical physics. The only relationship between quantum phenomena and QM in these interpretations is defined by the fact that QM predicts, in general probabilistically, the outcomes of quantum experiments. The probabilistic (or statistical) nature of these predictions is in accord with what is experimentally observed because no other predictions concerning such outcomes are in general possible, as concerns kinematic or dynamical variables, such as the position or the momentum, or the direction of spin. Such quantities as mass, charge, or spin are invariant. (There are certain specific situations, such those of the EPR-type experiments, where exact predictions are ideally possible, bur with important qualifications, explained below.) These predictions are, moreover, only possible by using rules *added* to the formalism rather than being part of it, such as Born's rule, which relates (essentially, by using complex conjugation) complex quantities of the formalism to real numbers corresponding to the probabilities of quantum events. Arguments to the effect that such rules are inherent in the formalism have been offered, but they are not commonly accepted. The Heisenberg postulate remains the grounding postulate of all RWR interpretations.

The concept itself of reality-without-realism is based in more general concepts of reality and existence, assumed here to be primitive concepts and not given analytical definitions. By "reality" I refer to that which is assumed to exist, without making any claims concerning the *character* of this existence or reality, claims that, as explained below, define realism. The absence of such claims allows one to place this character beyond representation or even conception, which defines the RWR view. I understand existence as a capacity to have effects on the world. The assumption that something is real, including of the RWR-type, is made, by inference, on the basis of such effects. The RWR view is grounded in the assumption that observable (either immediately or via a mediation of observational instruments) effects of physical reality allow for a representation these effects but not necessarily a representation (the weak RWR view) or even a conception (the strong RWR view) of how these effects are possible. The latter representation or conception may not be possible and is not in RWR interpretation in the case of the ultimate reality responsible for quantum phenomena, making this reality invisible to thought. It follows this reality or nature or matter, in the first place, are assumed to exist independently in the first place, just as it would be in classical theories or relativity or other realist theories, where, however, all strata of nature or matter considered are assumed to be visible to thought. The assumption of the independent existence of nature or matter essentially amounts to the assumption that it has existed before we existed and will continue to exist when we will no longer exist. Even this assumption, which still belongs to thought, has been challenged, even to the point of denying that the ultimate nature of reality is material rather than mental. Plato is the most famous ancient and Bishop Berkeley as the most famous modern case of this questioning. Such views are useful in suggesting that any conception of how anything exists, or even that it exists, including when assumed to be unavailable to human thought, still belongs to thought. It need not follow, however, that something which such concepts represent or to which they relate otherwise, possibly placing it beyond representation or even conception, does not exist.

Quantum phenomena would not be possible without our interaction with nature by means of experimental technology and our specific (human) ways of observing phenomena and thinking about them, which makes them visible to thought or even to our immediate phenomenal perception or consciousness. (Not all perceptions or forms of thought are conscious.) In RWR interpretations, nature has no quantum objects; and when it comes to its ultimate workings, at least those responsible for quantum phenomena, nature is beyond knowledge or, in strong RWR interpretations, conception. Hence, the term "workings," "nature," or "existence" would not ultimately apply either, any more than any other terms or concepts. Modern physics gave us new, mathematical-experimental, means of dealing with the world by interacting with it by means of experimental technology and mathematics (as a form of thought). In the present view, however, it gave us no more than such means, even in classical physics or relativity, where the assumption that the theory actually (ideally) represent nature is workable for all practical purposes. In any RWR interpretation, the concept of a quantum object is an idealization created in response to our interactions with nature by means experimental technology resulting in quantum phenomena. The present interpretation goes further by assuming the Dirac postulate, which makes the concept of quantum object an idealization (of the RWR type) only applicable at the time of observation.

Importantly, however, the present (strong RWR) interpretation does not a uniform or otherwise unified character of the ultimate, RWR-type, reality considered in QM or QFT, a character only manifesting itself differently in quantum experiments. This assumption is in conflict with strong RWR interpretations, which preclude any conception of this reality and, hence, that of its unity or oneness, uniform or not. While each time unknowable or even unthinkable, invisible to thought, an RWR-type reality is assumed to be each time different. This is what makes each quantum phenomenon, as an effect of this reality, individual and unrepeatable, unique, manifesting the unique, but still inconceivable, nature of the reality ultimately responsible for it each time one encounters this reality through its effects. One can always repeat the setup of a given measurement, because this setup can be classically controlled. Not so, however, as concerns the outcome of this repeated measurements. Such outcomes are ideally the same and are ideally predictable exactly in in classical or relativistic experiments dealing with individual or simple systems, with probability only entering when these systems have a great mechanical complexity, as in classical statistical physics or chaos theory. By contrast, these outcomes will in general (apart from special cases) be different in identically prepared quantum experiments, no matter how elementary the quantum object considered. As explained below, while possible, even preparing a given state, say, that of a "spin-up," as manifested in the corresponding measurement, cannot in general be done in a single preparation, but only by post-selecting the required preparation.

A brief outline of realist thinking may help to sharpen the nature of the RWR view. Realist thinking is manifested in the corresponding theories, commonly representational in character. Such theories aim to represent the reality they consider, in modern, post-Galilean, physics primarily by mathematized models, suitably idealizing this reality. It is even possible to aim, including in quantum theory, for a strictly mathematical representation of this reality apart from physical concepts, at least as they are ordinarily understood, say, in classical physics or relativity. It is also possible to assume an independent structure (defined by properties and relationships among them) of the reality considered, while admitting that it is either (A) not possible to represent this architecture or (B) even to form a rigorously specified concept of it, either at a given moment in history or even ever. Under (A), a theory that is merely predictive could be accepted for lack of a realist alternative, usually with the hope that a future theory will do better by being a representational theory. Einstein and, often following him, others held this type of view of QM. What, then, grounds realism most fundamentally is the assumption that the ultimate constitution of reality possesses properties and the relationships between them, or, as in (ontic) structural realism, just a structure, the more elemental constituents of which are not defined in terms of properties [Ladyman 2016]. Such properties, relationships, or structures may either be ideally represented and hence known, or be unrepresentable or unknown or even unknowable, but are still conceivable, usually with a hope that they will be eventually so represented. Most realist theories are representational. In considering physics, the concept of realism just outlined is often called "scientific realism." However, this outline would apply to most forms of realism in science or philosophy. It does not cover *all* forms of realism, which would be impossible. I shall also refer, as is common, to realist theories as ontological.[10]

Thus, classical mechanics (used in dealing with individual objects and small systems, apart from chaotic ones), classical statistical mechanics (used in dealing, statistically, with large classical systems), chaos theory (used in dealing with classical systems that exhibit a highly nonlinear behavior), or relativity, special and general, are realist theories. While classical statistical mechanics does not expressly represent the overall behavior of the systems considered because their mechanical complexity prevents such a representation, it assumes that the individual constituents of these systems are represented by classical mechanics. In chaos theory, which, too, deals with systems consisting of large numbers of atoms, one assumes a mathematical representation of the behavior of these systems. Relativity posed major, even insurmountable, difficulties for our immediate spatiotemporal phenomenal intuition, because the relativistic law of addition of velocities (defined by the Lorentz transformation) in special relativity, $s = \frac{v+u}{1+(vu/c)^2}$, for collinear motion ($c$ is the speed of light in a vacuum), runs contrary to any possible intuitive conception. Our phenomenal intuition cannot conceive of, visualize, this kind of motion, thus, making this concept of motion no longer a mathematical refinement of a daily sense of motion as the concept of motion is in classical physics. Relativity was the first physical theory that defeated our ability to form a phenomenal visualization of an elementary individual physical process, although the concept of (classical) field in classical electromagnetism already posed certain complexities in this regard. Bohr did not miss this point: "I am glad to have the opportunity of emphasizing the great significance of Einstein's theory of relativity in recent development of physics with respect to our emancipation from the demands of visualization" [Bohr 1987, v. 1, p. 115-116]. Emancipation! This is not a casual word choice, rarely, if ever, found in Bohr. Special, as well as general, relativity, however, still offer mathematically idealized conceptual representations of the physical reality they consider, and in this respect allowed this reality still to be visible to thought.

---

[10] Although the terms "realist" and "ontological" sometimes designate more diverging concepts, these terms are commonly close in their meaning and will be used, as adjectives, interchangeably here. I shall adopt "realism," as a noun, as a more general term and refer by an "ontology" to the representation or conception of the reality considered by a given theory. Another, relatively common, term for realist theories, sometimes with additional specificities (not important for my argument here), is "ontic," coming, as does ontological, from the ancient Greek *on* (Being).

Quantum physics brought this emancipation to a more radical level, that of the invisible to thought, mathematically reflected in the Hilbert space (or analogous) formalism over $\mathbb{C}$. Event this mathematics itself poses difficulties of seeing this formalism as representing anything physical in space and time, represented in all physical theories thus far as concepts over $\mathbb{R}$, to which, the formalism relates, with the help of Born's rule which converts complex quantities into real one, by means of probabilistic predictions of the outcome of measurement. As Bohr noted in 1927 (before the Hilbert space version of the formalism was introduced, while the role of $\mathbb{C}$ in the formalism was already in place), "the symbolic [rather than representational] character of Schrödinger's method appears not only from the circumstance that its simplicity, similarly to that of the matrix theory [of Heisenberg], depends essentially upon the use of imaginary arithmetic quantities" [Bohr 1987, v.1, p. 76].

All theories just mentioned, apart from QM, are based in the assumption, defining all "epistemologically classical theories," as they may be called (which designation would apply to relativity as well), that one can observe the phenomena considered without disturbing them. As a result, these phenomena can be identified with the corresponding physical objects and their independent behavior and, ideally, represent this behavior and predict it, in the case of individual or simple systems, ideally exactly, by using this representation. This is no longer possible in dealing with quantum phenomena, regardless of interpretation, and hence also in realist interpretations of QM, or alternative theories, such as Bohmian mechanics, of quantum phenomena. On the other hand, this situation opens the possibility of RWR interpretations of QM or QFT. The irreducible role of measuring instruments in the constitution of quantum phenomena grounded Bohr's interpretation, *in all its versions*. As noted from the outset, Bohr adjusted, sometimes significantly, his interpretation.

As Bohr argued in the Como lecture, which presented his first interpretation of QM (but the argument was retained in all versions of his interpretation, including the ultimate, RWR, one), in classical physics and relativity "our … description of physical phenomena [is] based of the idea that the phenomena concerned may be observed *without disturbing them appreciably*" [Bohr 1987, v. 1, p. 53; emphasis added]. By contrast, "any observation of atomic phenomena will involve an *interaction [of the object under investigation] with the agency of observation* not to be neglected" [Bohr 1987, v. 1, p. 54; emphasis added]. One should keep in mind the subtle nature of this contrast: the interaction between the object under investigation and the agency of observation *gives rise* to a quantum phenomenon rather than *disturbs* it. Bohr became weary of the language of "disturbing of phenomena by observation" [Bohr 1987, v. 2, p. 64].[11] Bohr grounded his interpretation (in all its versions) in this role and, in the ultimate version of his interpretation, in the strong RWR concept of reality, as applied to quantum objects, placed beyond conception and thus made invisible to thought. The behavior of the observable parts of measuring instruments and, with them, quantum phenomena were idealized as representable by means of classical physics, by the Bohr postulate. Eventually, Bohr adopted the term "phenomenon" to refer strictly to what is observed in measuring instruments, as effects of their interaction with quantum objects:

> I advocated the application of the word phenomenon exclusively to refer to *the observations obtained under specified circumstances*, including an account of the whole experimental arrangement. In such terminology, the observational problem is free of any special intricacy since, in actual experiments, all observations are expressed by unambiguous statements referring, for instance, to the registration of the point at which an electron arrives at a photographic plate. Moreover, speaking in such a way is just suited to emphasize that the appropriate physical interpretation of the symbolic quantum-mechanical formalism amounts only to predictions, of determinate or statistical character, pertaining to individual phenomena appearing under conditions defined by classical physical concepts [describing the observable parts of measuring instruments]. [Bohr 1987, v. 2, p. 64; emphasis added]

---

[11] Relativity represented a step in this direction, insofar as, in contrast to Newtonian mechanics, space and time were no longer seen as preexisting (absolute) entities then measured by instruments, such as rods and clocks, but were instead defined by the latter in each local reference frame. Still the interference of observational instruments into the behavior of the objects considered could be disregarded, thus allowing, as in classical physics, identification of these objects with the observed phenomena for all practical purposes. Hence, the objects under investigation can be considered independently of their interactions with measuring instruments. Quantum phenomena preclude this type of idealization, to the point of, in RWR interpretations, making the ultimate nature of the reality responsible for quantum phenomena invisible to thought. Bohr often reflected on these affinities, as well as differences, between relativity and quantum theory (e.g., [Bohr 1987, v. 1, p. 115, v. 2, p. 41, 1935, pp. 701-702]).

As defined by "*the observations* [already] *obtained under specified circumstances*," phenomena refer to events that have already occurred and not to possible future events, such as those predicted by QM. This is the case even if these predictions are ideally exact or deterministic, which they can be in certain circumstances, such as those of EPR type experiments. The reason that such a prediction cannot define a quantum phenomenon is that a prediction for variable $Q$ (for example, that related to a coordinate, $q$) cannot, in general, be assumed to be confirmable by a future measurement, in the way they can be in classical physics or relativity. One can always perform a complementary measurement, that of $p$ (the momentum), which will leave any value predicted by using $Q$ undetermined by the uncertainty relations, which in principle preclude associating a physical reality corresponding to a coordinate $q$ when one measures $p$ [Plotnitsky 2021a, pp. 210-212]. As earlier, I use capital vs. small letters to differentiate, as is necessary, Hilbert-space elements, here operators, like $Q$ and $P$, associated with predicting the values of measured quantities, like $q$ and $p$, observed on measuring instruments. Hence, one can never speak of both variables unambiguously, even if they are associated with measuring instruments, while any references, even that to a single property of a quantum object considered independently is ambiguous. In classical physics, this difficulty does not arise because one can, at least in principle, always define both variables simultaneously and unambiguously speak of the reality associated with both variables and assign them to the object itself. By contrast, in any quantum experiment we always deal with a system containing an object and an instrument. Thus, in considering quantum phenomena, on the one hand, there is always a discrimination between an object and an instrument, and, on the other, the impossibility of physically separating them. This impossibility compelled Bohr to speak of "the essential ambiguity involved in a reference to physical attributes of objects when dealing with phenomena where no sharp distinction can be made between the behavior of the objects themselves and their interaction with the measuring instruments," as opposed to a reference to what is observed which, as classical by the Bohr postulate, can be unambiguous and communicated as such [Bohr 1987, v. 2, p. 61].

There are several reasons for adding the Dirac postulate to the Heisenberg and Bohr postulates, defining Bohr's ultimate interpretation, in QM and, as noted, even more so in QFT, as considered in [Plotnitsky 2021a, pp. 273-306, 2021b, 2022b], beginning with the fact that no properties can be assigned to a quantum object apart from observations in Bohr's interpretation. In fact, in any strong RWR interpretation, nothing at all can be said or even thought about what happens between observations. Need one, then, still speak of quantum objects between observation? Also, as explained in detail in Section 5 (although, as just indicated, this point is implied by Bohr's concept of a phenomenon), in quantum physics in each experimental arrangement defining an observation one must, regardless of interpretation, discriminate "between those parts of the physical system considered which are to be treated as measuring instruments and those which constitute the objects under investigation" [Bohr 1935, p. 701]. The difference between them is, however, not uniquely defined. This is related to the arbitrariness of the "cut," considered in Section 5. For the moment, it follows that it is how one sets up an experiment that defines what is "the object under investigation" in this experiment, which invites assuming the concept of a quantum object to be applicable only at the time of observation. The situation has additional complexity because, while still treatable by means of QM, "the object under investigation" in a quantum experiment may not be strictly a quantum object: it may be *partly* classical, for example, containing the cat of the cat experiment, which is why Bohr speaks here of "objects under investigation" rather than quantum objects. This object, however, can only be only partly classical because it must contain, as its part, a properly quantum object, such as an electron or a photon, or some composite quantum object, to observe quantum effects. I shall return to this aspect of the situation, which bears importantly on the cat experiment, in Section 5, merely noting here that, a classical part, such as the cat, of such a combined object is always the same object and hence does not obey the Dirac postulate, which only applies to quantum objects. In accordance with Bohr's concept of a phenomenon, whatever is the object of investigation in a quantum experiment, it cannot be considered independently of its interaction with the measuring instrument, thus, making any quantum experiment and hence quantum theory involve both a combination and a separation of an object and an instrument. Because, however, the object under investigation in a *quantum* experiment must contain a properly quantum object, this is also a combination of what is invisible to thought and cannot be communicated unambiguously, and what is visible to thought, via observational instruments, and can be communicated unambiguously. Hence, as noted, the Bohr postulate also manifests the transition, via observation, from the ultimate, "quantum," reality to the classical level of observation, and conversely, in the initial preparation of an experiment, from the classical level of observation to the ultimate, "quantum," reality. According to Bohr:

> The essential lesson of the analysis of measurements in quantum theory is thus the emphasis on the necessity, in the account of the phenomena, of taking the whole experimental arrangement into consideration, in complete conformity with the fact that all unambiguous interpretation of the quantum mechanical formalism involves the fixation of the external conditions, defining the initial state of the atomic system concerned and the character of the possible predictions as regards subsequent observable properties of that system. Any measurement in quantum theory can in fact only refer either to a fixation of the initial state or to the test of such predictions, and it is first the combination of measurements of both kinds which constitutes a well-defined phenomenon. [Bohr 1938, p. 101]

One begins an experiment by classically preparing an observational instrument and registering, at time $t_{prep}$, the data obtained by the interaction between this instrument and a quantum object, thus setting up the workings of the ultimate reality considered, placed beyond representation or even conception, by the Heisenberg postulate. This is the classical to the quantum, the visible to thought to the invisible to thought, conversion. Then by setting up a new observational device, one makes a new observation at time $t_{observ}$ registering an outcome of the experiment, possibly as predicted by QM, in which case the observational instrument needs to be prepared accordingly, for, as explained, one can always perform a different type of measurement at this moment in time. This is the quantum to the classical, the invisible to thought to the visible to thought, conversion.

If, however, in assuming, as I do here, the Dirac postulate, a quantum object is only an idealization defined by an observation, rather than of something that exist independently (vis-à-vis the ultimate reality responsible for quantum phenomena assumed to exist independently), could one still speak of the same quantum object, say, the same electron, in two successive observations, with the second confirming the prediction based on the first of the formalism of QM? The case can be given a strictly RWR interpretation, insofar as all these properties are, physically, those of measuring devices, impacted by quantum objects, rather than of these objects themselves, placed beyond representation or conception. Rigorously speaking, if the concept of a quantum object is only applicable at the time of observation, then a prediction based on a given measurement and the new measurement based on this prediction could only concern a new quantum object, and not an object that one measured earlier in making a prediction. Accordingly, one deals with two different quantum objects, two different electrons, for example. To consider them as the same electron is, however, a permissible idealization in low-energy QM, or low-energy QFT, regimes. By contrast, speaking of the same electron in successive measurements in high-energy (QFT) regimes is meaningless, because these measurements can register quantum objects of different types, say, in the case quantum electrodynamics (QED) an electron in the initial and a positron or photon in the next measurement [Plotnitsky 2021a, pp. 279-292, 2021b]. QFT supports adding the Dirac postulate to the Heisenberg and Bohr postulate, in RWR interpretations, but, as I argue here, there are reasons also to do so in low-energy (QM) regimes, including, it may be shown, the complexities involved in the double-slit and related experiments [Plotnitsky 2022b].

On the other hand, there is no difficulty in speaking of the same classical object, even if it is part of the object of investigation by quantum means, such QM, in a quantum experiment, which, again, require a properly quantum object, such as emitted particle in the cat experiment, to be a quantum experiment. At all stages of the cat experiment, we deal with the same cat, dead or alive. The state of a classical object can change in time, just as the state of a measuring instruments does when impacted by a quantum object. If one tosses a coin, its state will change throughout its trajectory before it falls with either head or tail side up. Of course, this "sameness" is an idealization, possible and necessary in classical physics or relativity, or in dealing with classical objects, including measuring instruments, in quantum physics. As Heraclitus famously said, one cannot step in the same river twice because neither the river nor the one who steps into it is the same. Such concepts, however, do not apply to quantum objects, such as electrons, which, while they can change their location, momentum, or energy, are considered as strictly indistinguishable from each other in terms of any invariant characteristics, such as mass, changes, or spin. In RWR interpretations, these quantities are still only observable as effects of the interactions between quantum objects and measuring instruments, because no concepts apply to quantum objects, whether one defines them as existing independently, as in Bohr, only at the time of measurement, as here.

Two key concepts defining classical physics and relativity, (classical) "measurement" and (classical) "causality," become no longer applicable in quantum theory in RWR interpretations. The term "measurement" is a remnant of classical physics and the history that shaped it, beginning with ancient Greek thinking and the rise of geometry, geo-*metry*, there. In Bohr's and the present view, a quantum measurement *does not measure* or, in the first place, *is not an observation of* any property of the ultimate constitution of the reality responsible for quantum phenomena, a property that this reality would be assumed to possess before or even during the act of observation. The concept of observation requires a redefinition as well. An act of observation in quantum physics establishes, *creates,* quantum phenomena by an *interaction* between the instrument and the quantum object. This act is a unique event of creation.[12] This view also gives a new meaning to and gives a central significance to the category of event, as defining a new physical situation each time, akin to important events that transform the situation in life or culture, including politics, except that in quantum physics every event of observation radically transforms the situation and redefines the possibly future vis-à-vis the preceding events, no longer meaningful for predictions concerning the future from this point on. As a result, quantum theory becomes a theory of transition probabilities between events, thus defined by experimental technology and our decisions concerning which experiment to performed. Then what is so observed as the data or information can be measured classically, just as one measures what is observed in classical physics. There, however, what is observed or measured could be associated with the object considered. In quantum physics, there is a difference between observations, which construct phenomena, and measurements, which classically measure physical properties of the phenomena thus constructed. In speaking of "quantum measurement," I refer to this whole process. It follows that measuring instruments must contain both the visible (even to our immediate phenomenal perception) observable classical stratus and the quantum stratum, which enables their interactions with quantum objects. This interaction is quantum and cannot be observed and, in RWR interpretations, be visible to thought. It is, in Bohr's language, "irreversibly amplified" to the classical level of observable effects, such as a spot left on a silver screen [Bohr 1987, v. 2, p. 73].[13]

---

[12] While this formulation echoes J. A. Wheeler's invocation, inspired by Bohr, of "an elementary act of creation," the present view and, I would argue that of Bohr, may be different from Wheeler's view of a "participatory universe" [Wheeler 1983, pp. 189, 194; Wheeler 1981]. As existing independently of us, as it is assumed to be in the present and Bohr's views, (the reality of) the universe is not participatory; only independent phenomena, as created by us, and the world we experience are participatory. It is possible that by "a participatory universe," Wheeler refers to the world of our experience, including of quantum phenomena as our acts of creation. Wheeler, however, does not qualify his view in this way, and in any event, he never advances, and does not appear to adopt, the idea of reality without realism, which underlies the present and, I argue, Bohr's view, even though Bohr does not use the term.

The nature of causality in QM changes as well, as classical causality is no longer possible in RWR interpretations. By "classical causality" I refer to the claim that the state, $X$, of a physical system is determined, in accordance with a law, at all future moments of time once its state, $A$, is determined at a given moment of time, and state $A$ is determined by the same law by any of the system's previous states. This assumption implies a concept of reality, which defines this law, thus making this concept of causality ontological or realist. There are several reasons for my choice of "classical causality," rather than just causality, used more commonly for this type of concepts. The main one is that it is possible to introduce alternative, probabilistic, concepts of causality, applicable in QM, including in RWR interpretations, where classical causality does not apply (e.g., [Plotnitsky 2021a, pp. 207-218]). Some, beginning with P. S. Laplace, have used "determinism" to designate classical causality. I define "determinism" as an epistemological category referring to the possibility of predicting the outcomes of classically causal processes ideally exactly. In classical mechanics, when dealing with individual or small systems, both concepts become equivalent. On the other hand, classical statistical mechanics or chaos theory are classically causal but not deterministic in view of the complexity of the systems considered, which limit us to probabilistic or statistical predictions concerning their behavior.

In quantum phenomena, deterministic predictions are not possible even in considering the most elementary quantum systems. This is because the repetition of identically prepared quantum experiments in general leads to "different recordings" of the observed data (associated with the kinematic and dynamical variables), and unlike in classical physics, this difference cannot be diminished beyond the limit, defined by $h$, by improving the capacity of our measuring instruments [Bohr 1987, v. 2, p. 73]. "Recordings" refers to both those of the initial measurement, enabling a prediction, and those of the second measurement, which would verify this prediction, a combination that, as explained above, generally defines an experiment in physics, but takes a new meaning in quantum physics. These recordings will be different either one repeats the whole procedure in the same set of experimental arrangements or if one builds a copy of the apparatus and sets it up in the same way, as we do to separately verify the outcomes of experiments. Either repetition is always possible because the preparations of the instruments could be controlled classically. On the other hand, their interaction with quantum objects (or in the present view, the ultimate reality responsible for quantum phenomena and, at the time of measurement, quantum objects) cannot be controlled, which compelled Bohr to speak of "the finite and uncontrollable interaction between the object and the measuring instruments in the field of quantum theory" [Bohr 1935, p. 700]. The respective probabilities of the first and the second measurements are independent of each other. The most crucial, however, is the difference in the outcomes of the second (predicted) measurement in repeated setups. As noted, one can prepare any given state, say, that of a "spin-up," as manifested in the corresponding measurement, even though one cannot do so in a single experimental preparation but only by post-selecting the required preparation. By contrast, the outcome of the second (predicted) measurement cannot be controlled at all, only allowing one to predict the probability or, if the experiment is repeated, statistics of the outcome.

---

[13] The physical nature of this "amplification" is part of the problem, commonly, including by this author, seen as unsolved (although there are claims to the contrary, for example, on lines of decoherence or consistent histories approaches), of the transition from the quantum to the classical. The subject is beyond my scope here. Fortunately, quantum phenomena and QM allow us to bypass this problem in quantum measurements or predictions, seen here as the transition from the invisible to thought to the visible to thought (or vice versa in a preparation). As Bohr noted, QM is "justified only by the possibility of disregarding in its domain of application the atomic structure of measuring instruments themselves in the interpretation of the results of experiments" [Bohr 1937, p. 88]. This disregard, as Bohr observed, may lead to new complexities in high-energy physics and QED. As he said, invoking, again, a renunciation of visualization: "For a correlation of still deeper laws of nature involving not only the mutual interaction of the so-called elementary constituents of nature but also the stability of their existence, this last assumption can no longer be maintained, as we must be prepared for a more comprehensive generalization of the complementary mode of description which will demand a still more radical renunciation of the so-called visualizations" [Bohr 1937, p. 88]. As it happens, QFT (in high-energy regimes) still disregards "the atomic structure of measuring instruments," which may be responsible for the appearance of infinities and the necessity of renormalization and other, still unresolved, complexities there.

The statistics of the outcomes of multiply repeated experiments performed in both such experimental settings will be the same. On the other hand, an individual quantum experiment cannot be reproduced, as. is always possible to do so in classical physics, because the interference of measurement can be neglected or controlled, at least in principle. All data observed in quantum experiments remains classical, by the Bohr postulate, and hence visible to thought (or even to the immediate phenomenal perception) and can be communicated unambiguously. Unlike in classical physics, however, this data cannot be recreated by a different system, which combines a quantum object (in the present view, again, a concept only applicable at the time of observation) and an apparatus, the observable part of which is described classically. This situation embodies the no cloning theorem [Park 1970, Dieks 1982, Wootters and Zurek 1982].

As noted, the probabilistic or statistical character of quantum predictions must, on experimental grounds, hold in interpretations of QM or alternative theories of quantum phenomena (such as Bohmian mechanics) that are classically causal. QM or QFT, in RWR interpretations, are not classically causal because the ultimate nature of reality responsible for quantum phenomena is assumed to be beyond a representation or conception. Classical causality would imply at least a partial conception and even representation of this reality. These circumstances imply a different reason for the recourse to probability in quantum theory in RWR interpretations. According to Bohr:

> [I]t is most important to realize that the recourse to probability laws under such circumstances is essentially different in aim from the familiar application of statistical considerations as practical means of accounting for the properties of mechanical systems of great structural complexity. In fact, in quantum physics we are presented not with intricacies of this kind, but with the inability of the classical frame of concepts to comprise the peculiar feature of indivisibility, or "individuality," characterizing the elementary processes. [Bohr 1987, v. 2, p. 34]

The "indivisibility" refers to the indivisibility of phenomena in Bohr's sense, defined by the impossibility of considering quantum objects independently from their interactions with these instruments. "Individuality" refers to the assumption that each phenomenon is individual and unrepeatable, as well as discrete relative to any other phenomenon, and correlatively, to the essential randomness of individual quantum phenomena. Collectively they may not be strictly random by virtue of one or another form of quantum correlations (such as EPR-type correlations, at stake in Bell's or Kochen-Specker theorem), which are, however, strictly quantum as well and not found in classical phenomena. This randomness is not found in classical physics, because even when one must use probability there, at bottom one deals with individual process that are classically causal and in fact deterministic. Hence, in classical physics, randomness does not ultimately exist or is assumed ultimately not to exist; only probability does. In principle, one can isolate an individual constituent of the structurally complex mechanical system, say, a molecule of a gas, something that, as classical, is in its behavior, visible to thought, and predict its behavior ideally exactly. It is, however, the "in principle" that is crucial, because this is never possible in considering individual quantum systems, no matter how elementary. By the same token, such systems or (since the term "system" is not ultimately applicable either, except at the time of measurement) the ultimate nature of the reality considered can never be made visible to thought, which is, again, the reason why they cannot be assumed to be classically causal or predicted deterministically. In fact, as explained, the possibility of never observing quantum objects as isolated defined Bohr's concept of a phenomenon, and in the present (more radical) view, quantum objects are only defined, still as invisible to thought, at the time of observation. Quantum physics, then, *contains* an essential randomness not found in classical physics, which is at bottom classically causal and, when it comes to the behavior of its elemental individual constituents, deterministic, thus making the recourse to probability a practical, epistemological matter, as Bohr says. A coin toss is an example of a classical probabilistic system. The outcome can, *in* *practice*, only be predicted probabilistically due to the mechanical complexity of the process, beginning with the motion of the hand tossing it. However, this is still a classically causal process, the outcome of which is determined and could, *in principle*, be predicted ideally exactly with sufficient technical and computational capacities, which is, as explained, the meaning of "classically causal." The recourse to probability is practical, epistemological, due to our lack of knowledge concerning the underlying behavior of the systems considered. In the case of any quantum system, no matter how simple, this idealization is not possible: its behavior is not assumed to be classically causal in RWR interpretations, and as invisible to thought, it cannot be so assumed.[14] Quantum physics, however, only *contains* this randomness, rather than is entirely random, because it allows for probabilistic or statistical predictions (purely random events do not, which makes it impossible to handle them scientifically) and, more crucially, correlations. One of the greatest mysteries of quantum phenomena is how random individual events can, under certain circumstances, give rise to an order, even if only a (statistical) correlational order [Plotnitsky 2021a, pp. 253-256]. QM predicts these correlations, but at least in RWR interpretations, it does not explain them, any more than it explains how any single outcome of an observation or measurement, comes about. The emergence of either is invisible to thought.

I shall now explain Bohr's concept of complementarity, especially, as it appears in his ultimate, strong RWR interpretation, where it applies to phenomena in Bohr's sense as outlined above. As defined generally complementarity is characterized by:

(A) a mutual exclusivity of certain phenomena, entities, or conceptions; and yet
(B) the possibility of considering each one of them separately at any given point; and

---

[14] One might further distinguish between indeterminacy, as a more general category, and randomness, as a most radical form of indeterminacy, when a probability cannot be assigned to a possible event, which may also occur unexpectedly. Both indeterminacy and randomness only refer to possible future events and define our expectations concerning them. Once an event has occurred, it is determined. An indeterminate nature of events may either allow for assuming an underlying classically causal architecture (which may be temporal) of the physical reality responsible for this nature, whether this process is accessible to us or not, or disallow for making such an assumption. The first case, as just explained, defines indeterminacy in classical physics, such as classical statistical physics or chaos theory, or more radically in considering the so-called algorithmic complexity, such as Kolmogorov complexity (also known as Solomonoff-Kolmogorov-Chaitin complexity), which may not be computable, but still for practical, epistemological reasons. The second is found in QM or QFT in RWR interpretations. According to Bohr, the idea of indeterminacy (or, again, randomness) apart from a classically causal order has "hardly been seriously questioned until Planck's discovery of the quantum of action" (Bohr 1938, p. 94). As he said on a later occasion (in 1949): "[E]ven in the great epoch of critical [i.e., post-Kantian] philosophy in the former century, there was only a question to what extent a priori arguments could be given for the adequacy of space-time coordination and causal connection of experience, but never a question of rational generalizations or inherent limitations of such categories of human thinking" (Bohr 1987, v. 2, p. 65). Even more radical philosophical questionings of the classical idea or ideal of causality, such as those by David Hume, are those of our epistemological capacity to perceive the underlying classically causal world, which would be presupposed at the ultimate level as inaccessible to us. It is impossible to ascertain that an apparently random sequence of events, events that occurred apparently randomly, was in fact random, rather than connected by some rule, such as that defined by classical causality, and there is no mathematical proof that any "random" sequence is actually random (e.g., Aaronson 2013, pp. 71-92). The sequences of indeterminate events that allow for probabilistic predictions concerning them is a different matter, although there is still no guarantee that such sequences are not ultimately underlain by classically causal connections in the case of quantum phenomena. Experimentally, again, quantum phenomena only preclude determinism, because identically prepared quantum experiments in general lead to different outcomes. It follows that the claim of quantum randomness can, in principle, be falsified, but establishing a classically causal theory or algorithm that reproduces the indeterminate or random data in question, which becomes no longer indeterminate random. This would imply that RWR interpretations, which precludes such connections, does not correspond to the ultimate nature of reality responsible for quantum phenomena. See [D'Ariano 1922], which establishes the existence of a falsifiable quantum random generator. In the present view, such a generator cannot be classical, because all classical (or relativistic) theories of individual systems are deterministic, that is, can be so idealized as such.

(C) the necessity of considering all of them at different moments of time for a comprehensive account of the totality of phenomena that one must consider in quantum physics.

The concept was never given by Bohr a single definition of this type. However, this definition may be surmised from several of Bohr's statements, such as: "Evidence obtained under different experimental conditions cannot be comprehended within a single picture, but must be regarded as complementary in the sense that only the totality of the phenomena [some of which are mutually exclusive] exhaust the *possible* information about the objects" (e.g., [Bohr 1987, v. 2, p. 40]). In classical mechanics, we can comprehend all the information about each object within a single picture because the interference of measurement can be neglected. This allows us to identify the phenomenon with the object under investigation and establish the quantities defining this information, such as its position and momentum, *in the same experiment*. In quantum physics, this interference cannot be neglected and leads to different, in fact mutually exclusive, experimental conditions for each measurement and their complementarity, in correspondence with the uncertainty relations. The situation implies two incompatible pictures of what is observed, as phenomena, in measuring instruments. Hence, the *possible* information about a quantum object, the information *to be found* in measuring instruments, could only be exhausted by the mutually incompatible evidence obtained under different experimental conditions. On the other hand, once made, either measurement, say, that of the position, will provide the *complete actual* information (manifested in measuring instruments) about the object, as complete as possible, at this moment in time. One could never obtain the complementary information, provided by the momentum measurement, at this moment in time, because to do so one would need simultaneously to perform a complementary experiment on it, which is impossible.

Thus, parts (B) and (C) of the above definition of complementarity are as important as part (A) and disregarding them can lead to a misunderstanding of Bohr's concept, often misleadingly identified with just (A). Bohr's complementarity is not only about a mutual exclusivity of things, but also about performing quantum experiments by human agents, in which a mutual exclusivity becomes necessary. That we have a free (or at least sufficiently free) choice as concerns what kind of experiment we want to perform is in accordance with the very idea of experimentation in science, including in classical physics [Bohr 1935, p. 699]. However, contrary to the case of classical physics or relativity, implementing our decision concerning what we want to do will allow us to make only certain types of predictions and will irrevocably exclude certain other, *complementary*, types of possible predictions. In other words, we have a freedom, at least a sufficient degree of freedom, of choice which experiment to perform in classical and quantum physics alike. In classical physics (or relativity), however, it does not matter in fundamental terms because all variables necessary for defining the future course of reality, in accord with classical causality, can always be determined at any moment in time, as there is no complementarity or the uncertainty relations. By contrast, by virtue of complementarity, it does matter in quantum physics: By staging, by decision, our experiments in one complementarity way or the other, we define the course of reality, even if only probabilistically, because, while we can control the set-up of the experiment, we cannot control the outcome. Such uncontrollable outcomes are no longer a matter of surprise that nature confronts us with but is instead what we expect from nature, or our interaction with nature, in quantum experiments. It also follows that we always, at any point, have a freedom, in any event, a sufficient degree of freedom to make this choice or to change our choice and thus a future course of reality. Beyond, as discussed below, the Bohr-EPR debate concerning the EPR experiment, this aspect of complementarity is related in Bell's and the Kochen-Specker theorem, or the Conway-Kochen free will theorem. The latter connections are, however, a separate subject beyond my scope here.

For the moment, more immediately complementarity is a reflection of the fact that, in a radical departure from classical physics or relativity, the behavior of quantum objects of the same type, say, electrons, or, again, the ultimate nature of reality responsible for quantum phenomena defined by such objects, is not governed by the same physical law, especially a representational physical law, in all possible contexts, specifically in complementary contexts. This leads to incompatible observable physical effects in complementary contexts. On the other hand, the mathematical formalism of QM offers correct probabilistic or statistical predictions of quantum phenomena *in all contexts*, in RWR interpretations under the assumption that the ultimate nature of reality responsible for quantum phenomena is invisible to thought.[15] However, as Bohr observed, reiterating his argument concerning the nature of quantum probability considered above:

> Just in this last respect [of the renunciation in each experimental arrangement of the one or the other of two aspects of the description of the physical phenomena] any comparison between quantum mechanics and ordinary statistical mechanics,—however useful it may be for the formal presentation of the theory,—is essentially irrelevant. Indeed we have in each experimental arrangement suited for the study of proper quantum phenomena not merely to do with an ignorance of the value of certain physical quantities, but with the impossibility of defining these quantities in an unambiguous way. [Bohr 1935, p. 699]

It might be noted that wave-particle complementarity, with which the concept of complementarity is often associated, had not played a significant, if any, role in Bohr's thinking, especially after the Como lecture. Bohr was always aware of the difficulties of applying the concept of physical waves to quantum objects or assuming both types of behavior, particle-like and wave-like, pertain to the same individual entities, such as each photon or electron itself, considered independently. Bohr's ultimate solution to the dilemma of whether quantum objects are particles or waves was that they were neither, any more than anything else, by the Heisenberg postulate. Instead, either "picture" refers to one of the two mutually exclusive sets of discrete individual effects, described classically by the Bohr postulate, of the interactions between quantum objects and measuring instruments, particle-like, which may be individual or collective, or wave-like, which are always collective, composed of discrete individual effects. An example of the latter are interference effects, composed of a large number of discrete traces of the collisions between the quantum objects and the screen in the double-slit experiment in the corresponding setup (when both slits are open and there are no means to know through which slit each object has passed). These two sets of effects may be seen as complementary, also when it comes to calculating the probabilities or statistics for each set of events, or, if one takes a Bayesian view, for each event of each set. The two types of effects involved are mutually exclusive and require mutually exclusive experimental setups to be observed. In classical physics, wave-like (radiation) and particle-like *objects* or (as they can be identified) phenomena were treated by two mutually exclusive theories, which is not the same as being complementary in Bohr's sense. The latter must include (B) and (C) part of the concept, applicable to the same (quantum) objects or the ultimate reality responsible for quantum phenomena, but leading two different phenomena by (A), depending on which setup one decided to use, predicted, differently, by the same theory, QM or QFT.

I would like, in closing my discussion of the (strong) RWR view, as defined by the role of the invisible to thought in quantum physics, briefly to reflect, from this perspective, on the EPR experiment and the Bohr-EPR exchange concerning it. My reflections follow [Plotnitsky 2021a, pp. 227-272], which offers a detailed discussion, although the angle of visible and invisible to thought is new. The case is, however, both exemplary and highly significant in this context, as Bohr, as noted above, realized in stressing the significance of the EPR experiment as "suited to emphasize how far, in quantum theory, we are beyond the reach of pictorial visualization" (Bohr 1987, v. 2, p. 59). One might give a new angle on and amplify Bohr's point, by arguing that that in their argument, EPR in effect assume that the independent reality of quantum objects is visible to thought. EPR's argument is, however, based on disregarding or at least not adequately considered the constitutive role of observational instruments in defining quantum phenomena in the way Bohr argued to be necessary, based on an analysis of this role in his reply [Bohr 1935]. In fact, while EPR do, unavoidably, refer to "measurement," EPR do not considered or even mention measuring instruments, the constitutive role of which in defining all physical variable concerned would make it difficult or even impossible to assume that the ultimate nature of reality responsible for quantum phenomena can be visible to thought.

---

[15] This situation is also responsible for what is known as "contextuality," which was considered from the RWR perspective in [Plotnitsky 2019, Plotnitsky 2021a], and, along different lines, in [Jaeger 2019, Howard 2021]. See also Khrennikov's extended survey [Khrennikov 2022].

EPR advanced the following argument based on the criterion of reality they which, they thought, equally applicable in classical and quantum physics: "*If, without in any way disturbing a system, we can predict with certainty (i.e., with probability equal to unity) the value of a physical quantity, then there exists an element of physical reality corresponding to this physical quantity*" [Einstein et al, p. 138]. While, however, this criterion is unproblematically applicable in classical physics, it is, Bohr contended, "ambiguous" in the case of quantum phenomena, because of the role of measuring instruments in defining all such quantities. EPR's argued that it is possible to ascertain "an element of reality" pertaining to a quantum object, the second, $S_2$, object of the EPR pair ($S_1$, $S_2$), independently of any interaction between $S_2$ and a measuring instrument (thus "without in any way disturbing the system"). This association is made possible by a prediction by means of QM, say, of variable $q$ (like that associated with the position operator $Q$) with "probability equal to unity," a prediction based on the measurement performed on $S_1$, as is indeed possible, at least ideally or in principle. As earlier, I use capital letters, $Q$ or $P$, to refer to the operators in the Hilbert space considered, and small letters, $q$ or $p$, to physical variables probabilistically predicted by the formalism by using $Q$ or $P$, which have no physical connections to $q$ or $p$ apart from these predictions in RWR interpretations. EPR argue that because this prediction, "with probability equal to unity," is possible "without in any way disturbing" $S_2$, this property could be ascertained as an element of physical reality pertaining to $S_2$, in accordance with their criterion. As such, it is in effect assumed to be visible to thought and unambiguously communicable, even though it is not actually observed or measured, and as such is not available to our immediate phenomenal perception at the time of prediction, or at any time, unless a measurement is performed, or even then because we can only perceive what is observed in measuring. While in the latter case it is in principle possible to associate such a measured quantity with an element of reality pertaining to the object itself, one is, obviously, outside the situation covered by EPR's criterion, because we no longer deal with a prediction and of course disturb the object by an observation. Hence, it is the concept of visible to thought that is crucial here.

Bohr counterargued that, while EPR's claim would work in classical physics, the situation was different in considering quantum phenomena, including those of the EPR type, because of the essential role of observational instruments in the constitution of all quantum phenomena and, thus, in any unambiguous application of the concept of reality or of an element of reality in quantum physics. This role, he argued, must be taken into consideration even in the case of predictions "with probability equal to unity" without a measurement previously performed on the system considered, $S_2$, and instead by using a measurement performed on $S_1$, as is ideally possible in the EPR case. As, however, I noted earlier and as Bohr argued in his reply, this prediction is not sufficient for assigning an element of physical reality to $S_2$, contrary to EPR's claim based on their criterion of reality, assumed by then to equally apply in both classical and quantum theory. This is, however, not the case. In classical physics, where one can, in principle, always measure and define both variables *simultaneously*, by neglecting the interference of observational instruments, it is possible to speak, at any moment in time, unambiguously of the reality in terms of its physical elements, thus visible to thought, associated with both conjugate classical variables, $Q$ and $P$ (as functions of real variables) and define them as pertaining to the object itself considered. Everything, at any point, is always visible to thought.

Not so, in quantum physics. Let us assume that by measuring $q_{S1}$ on $S_1$ and using the formalism, applied to $Q$ (an operator in a complex Hilbert space), one makes a prediction, "with probability equal to unity," concerning $q_{S2}$ associated, via a measuring instrument, with $S_2$ at some future time, $t$. If one measures $q$ at time $t$, one then will indeed obtain the value $q_{S2}$. However, one can, instead of $q$, always measure at time $t$ the complementary variable, $p$ (which would relate to the momentum operator $P$ in the formalism, although one does not use in a measurement). If one does so, the value of $q$ becomes completely undetermined, ambiguous, by the uncertainty relations. Hence, this measurement of $p$ would preclude associating any physical reality with the predicted value $q_{S2}$. Thus, $q_{S2}$, as defined by this prediction, may be visible to thought, but it can no longer correspond to any element of physical reality that can be associated with $S_2$, or at least there is no way to experimentally ascertain such a correspondence. $S_2$ is assumed to exist and hence be real, but to assume so is not the same as associating an element of reality with it and thus making it visible to thought.[16] This association is only possible if the measurement, confirming the predicted value, $q_{S2}$, is performed. Doing so, however, can be in principle precluded by making a complementary measurement and, thus, in contrast to classical physics (where both conjugate variables can always be assigned, corresponding to elements of reality, simultaneously), disabling the association of the predicted value $q_{S2}$ with $S_2$. This is so even if one assumes that one can associate an element of reality with $S_2$ as such, rather than only with a classical observed part of a measuring instrument, at the time of measurement. In other words, unless the corresponding measurement is performed, $q_{S2}$ can correspond to no elements of physical reality, and the possibility of establishing such a correspondence can be denied if one measures $p$ instead. This situation is captured by A. Peres's statement that "unperformed experiments have no result" [Peres 1978]. Bohr's claim concerning "an essential ambiguity" of EPR's criterion is defined by this situation, not considered by EPR in advancing this criterion, or by Einstein in his related arguments, based in the same criterion. As Bohr stated in the passage of his reply cited above, in view of complementarity, "we have in each experimental arrangement suited for the study of proper quantum phenomena not merely to do with an ignorance of the value of certain physical quantities, but with the impossibility of defining these quantities in an unambiguous way" (Bohr 1935, p. 699; also Bohr 1987, v. 2, p. 62). There is, he argues, absolutely no possibility to unambiguously define both "elements of reality" in question for $S_2$, "without in any way disturbing" it. One can only do so for one or other of complementarity quantities, say, $q$, by making the corresponding measurement on $S_1$, $q_{S1}$, and *predicting* the reality of the same type of element, $q_{S2}$, for $S_2$, still under the assumption that one could in principle perform the corresponding measurement. That, however, irrevocably precludes one from predicting the complementary element of reality, $p_{S2}$, for $S_2$, because any measurement of $p$ on $S_1$ was precluded by measuring $q_{S1}$. On the other hand, if one instead measures $p$ on $S_2$, which of course would require *disturbing* $S_2$, one in turn irrevocably precludes ascertaining $q_{S2}$ as an element of reality pertaining to $S_2$. Locating this ambiguity enables Bohr to argue that QM can be seen as *both* complete within its proper scope (as complete as nature allows our theory of low-energy quantum phenomena to be) *and* local, insofar it does not entail any physical action at a distance, or at least that EPR, who argued that QM is *either* incomplete *or* nonlocal in this sense, did not demonstrate otherwise, as explained in detail in [Plotnitsky 2021a, pp. 227-272].[17]

    The EPR-Bohr exchange was crucial for the development of Bohr's thinking, leading him to his ultimate, strong RWR interpretation and, correlatively, a deeper understanding of the nature of complementarity as a physical concept. It compelled Bohr eventually to adopt the view that no measurable quantity, even a single such quantity (rather than only both complementary quantities, as precluded by the uncertainty relations) and hence no element of reality can be attributed to a quantum object even at the time of measurement. While a quantum object was assumed by Bohr to exist and hence be real independently of observation, any reference to the nature of its reality becomes ambiguous, making Bohr speak of "the essential ambiguity involved in a reference to physical attributes of objects when dealing with phenomena where no sharp distinction can be made between the behavior of the objects themselves and their interaction with the measuring instruments" [Bohr 1987, v. 2, p. 61]. Such

---

[16] It should be kept in mind that, as Schrödinger was the first to note in defining entanglement, in dealing with entangled systems it is not possible to speak of the properties on each system separately or even (Schrödinger does not appear to go that far, at least not expressly) even of two separate systems [Schrödinger 1935, pp. 160-161]. However, once a measurement on $S_1$ is performed (thus also establishing it as quantum object], $S_1$ and $S_2$ are no longer entangled. This situation, it might be added, gives another justification to using the Dirac postulate, which only defines either system as such at the time of measurement.

[17] EPR were aware and assumed in their argument that both quantities cannot be measurement or predicted simultaneously, and their criterion of reality allows for assigning both "elements of reality" to $S_2$ without simultaneously predicting both. They argued, however, that the only alternative to their argument is the assumption of the nonlocality (an action at a distance) of QM or quantum phenomena [Einstein et al, p. 141]. Bohr counterargued that this nonlocality, as well as the incompleteness of QM, can be avoided by virtue of the ambiguity, and hence inapplicability, of EPR's criterion of reality to quantum phenomena, as here discusses. A detailed argument is offered in [Plotnitsky 2021a, pp. 227-272].

attributes, as elements of reality, can only be unambiguously ascribed (under the constraint of the uncertainty relations) to certain parts, *elements*, of quantum phenomena, defined by the observable parts of measuring instruments. This fact makes these elements open to being described by classical physics. While, however, Bohr associated the ultimate, invisible-to-thought, reality responsible for quantum phenomena with quantum objects, in the present interpretation, by the Dirac postulate, the concept of a quantum object is only applicable at the time of observation, still as an RWR concept, which precludes associating any attributes, *elements*, with it. The character of the ultimate reality considered as invisible to thought equally defines both interpretations arising from, and arguably reaching the most radical manifestation of, "the spirit of Copenhagen," in Heisenberg's memorable phrase "*der Kopenhagener Geist der Quantenheorie*," honoring Bohr's contribution to our understanding of quantum theory [Heisenberg 1930, p. iv)].

The existence, at least a possible existence, of a reality invisible to thought (the Heisenberg postulate), which is, at the same time, ultimately responsible for what is visible to thought in quantum phenomena (the Bohr postulate), is what Bohr saw as "an epistemological lesson of quantum mechanics" [Bohr 1987, v. 3, p. 12]. At least, this is an epistemological lesson of his *interpretation* of quantum mechanics, to which the present interpretation adds the Dirac postulate. Perhaps, however, quantum mechanics or physics in general cannot teach us lessons otherwise. It is just that there appears *now* (this has not always been the case) to be more consensus, albeit not an entirely unanimous one either, as concerns our interpretation of classical physics and relativity as realist theories. When it comes to QM or QFT, the proliferation of diverse (and sometimes incompatible) interpretations and the debate concerning them, still overshadowed by the Bohr-Einstein confrontation, continue with an undiminished intensity and no end in sight. But then, the stakes are high: the future of our understanding of nature and thought alike.

## 3. The Bohr-Schrödinger exchange on classical concepts in quantum measurement

Bohr's insistence, reflecting (in present terms) the Bohr postulate, on the indispensable role of classical physical concepts in considering measuring instruments is often misunderstood, and the subject is significant in the context of the cat experiment, which provides an additional reason for addressing this insistence in detail in this article. It is instructive to consider in this connection Schrödinger's comments on this aspect of Bohr's thinking in Schrödinger's letter after reading Bohr's reply to EPR (in a prepublication version), while working on his cat-paradox paper. The exchange, relevant to Schrödinger's overall argument in his paper, might have also affected his comments on the cat experiment, although the origin of the experiment appears to be a suggestion by Einstein [Fine and Ryckman 2020]. Schrödinger's (long) letter and Bohr's (brief) reply in part resume an earlier exchange, in 1928-1929, on the subject among Einstein, Schrödinger, and Bohr [Plotnitsky 2021a, pp. 32-34]. Schrödinger writes:

> You [Bohr] have repeatedly expressed your definite conviction that measurements must be described in terms of classical concepts. For example, on p. 61 of the volume published by Springer in 1931 [the original German edition of [Bohr 1987, v. 1]]: "It lies in the nature of physical observation, that all experience must ultimately be expressed in terms of classical concepts, neglecting the quantum of action" [Bohr 1987, v. 1, pp. 94-95]. And ibid. p. 74 "the invocation of classical ideas, necessitated by the very nature of measurement" [Bohr 1987, v. 1, p. 114]. And once again [in the reply to EPR] you talk about "the indispensable use of classical concepts in the interpretation of all [proper] measurement" [Bohr 1935, p. 701, where the printed version adds "proper"]. True enough, shortly thereafter you say: "The removal of any incompleteness in the present methods of atomic physics … might indeed only be affected by a still more radical departure from the methods of description of classical physics, involving the considerations of the atomic constitution of all measuring instruments, which it has hitherto been possible to disregard in quantum mechanics."
>
> This might sound as if what was earlier characterized as inherent in the very nature of any physical observation as an "indispensable necessity", would on the other hand after all just be a, fortunately still permissible, convenient way of conveying information, a way we presumably sometime will be forced to give up. If this were your opinion, then I would gladly agree. However, the subsequent stringent and clear comparison with the theory of relativity make me doubt whether, in what I just said, I have understood your view correctly. Because, if we considered the theory of relativity as a conceptual edifice in itself, without any relationships to quantum mechanics, we would presumably never be able to renounce the sharp separation between space and time *in any measurement*. Still, it seems possible that in connection with the unavoidable mutual modification of these two theories, *both* would be forced to shake off their classical eggshell—and that *this* is what you mean. (Letter to Bohr, October 13, 1935 [Bohr 1972-1996. v. 7, p. 505])

As Schrödinger admits ("it *seems* possible"), this may not be and, I would argue, is not what Bohr means. First, especially given that Bohr's argued in his reply to EPR that QM is a complete theory within its scope (as complete as nature allows our theory of nonrelativistic quantum phenomena to be), it is clear that "incompleteness" in Bohr's passage cited by Schrödinger does not refer to QM. It refers to the fact that at the time QFT was hardly adequately developed even in the case of QED. (H. Yukawa's meson theory of nuclear forces just introduced.) QED, too, only worked then to the first order of approximation, beyond which QED led to the appearance of infinities, which were only handled by renormalization fifteen years later. The passage in question was removed from Bohr in the published version of his response to the EPR paper [Bohr 1935], as Bohr explained in his reply to Schrödinger's letter. He said: "I have left out the reference to the possible significance of the atomic constitution of all measuring instruments for the solution of the still unexplained difficulties of electron theory [QED]. The reason is that together with Rosenfeld I am just about to finish a paper about a measuring problem in electron theory in which this question will be elucidated somewhat more fully" [Letter to Schrödinger, October 25, 1935, in Bohr 1972-1996, v. 7, p. 511].[18] Schrödinger was aware that Bohr referred to the incompleteness of QED and possibly QFT. It is clear, however, from this comment and related elaborations by Bohr including in [Bohr 1987, v. 1, pp. 89-91, 115], to which Schrödinger refers in his letter, that the point is not that the observable parts of measuring instruments should no longer be described by classical physics in QFT. Speaking of "a still more radical departure [than in QM] from the method of description of classical physics" only refers to a more radical situation in QFT as concerns a possible necessity, as against QM, of considering the atomic structure of measuring instrument, along with its observable parts, described classically. The latter aspect of quantum measurement would remain in place in QFT in Bohr's view, for the reasons discussed earlier in the present article and explained in Bohr's reply to Schrödinger's letter, while the atomic constitution interaction may need to be considered in a relativistic quantum theory. In fact, we still do not have a quantum theory that does so, and as currently constituted, QFT still works in the absence of such an account, which may be responsible for its difficulties.[19] It is not clear either whether such a theory is possible or necessary. QFT does contain unresolved difficulties (even apart from the absence of a quantum theory of gravity). It does work, however. In works well as a predictive theory or, one might argue, a framework, something sometimes referred to in theoretical physics as "phenomenology" (not to be confused with the used of the term in philosophy or when one speaks of our phenomenal representations). QED is now the best confirmed physical theory ever as concerns its predictions, probabilistic or statistical as they are, which predictions are, however, again what quantum experiments allows us, as things stand now.

Schrödinger, by contrast, appeared to think that Bohr believes that such a theory, as well as relativity, "would be forced to shake off their classical eggshells" of the description of measuring instruments, possibly even in QM. But, as I argue, this not what Bohr thinks: classical "eggshells" are part of phenomena, and unlike is classical physics, if one shakes them off or breaks them you will only create new eggs with eggshells, without ever exposing, making visible, what is inside. One cannot make an omelet out of the eggs of quantum phenomena, only new eggs. Any subdivision of a phenomenon can only result in a new phenomenon or phenomena, still each with classically described "shells," without ever exposing quantum objects. As Bohr explained later: "The individuality of the typical quantum effects finds its proper expression in the circumstance that any attempt of subdividing the phenomena will demands a change in the experimental arrangement introducing new possibilities of interaction between objects and measuring instruments which in principle cannot be controlled" [Bohr 1987, v. 2, p. 39]. Hence, Bohr speaks of closed phenomena, or the wholeness or indivisibility of phenomena. Bohr says in his letter to Schrödinger:

> However, these considerations [of the atomic structure of measuring instruments] do not have any close connection to the Einstein paradoxes and to the question of limitation of the [classically] causal description of quantum phenomena. On this point I must confess that I cannot share your doubts. My emphasis of [sic: on] the point that the classical description of experiments is unavoidable amounts merely to *the seemingly obvious fact that the description of any measuring arrangement must, in an essential manner, involve the arrangement of the instruments in space and their functioning in time, if we shall be able to state anything at all about phenomena*. The argument here is of course first and foremost that in order to serve as measuring instruments, they cannot be included in the realm of application proper to quantum mechanics. [Letter to Schrödinger, October 25, 1935, Bohr 1972-1996, v. 7, p. 511]

---

[18] This paper was not published and only became available in the same volume of Bohr's collected works [Bohr 1972-1996, v. 7, pp.195-209].
[19] See, note 13 above.

In other words, measuring instruments in their observable parts are and must be visible to thought and even to our immediate phenomenal perception, to "*be able to state anything at all about phenomena*" and thus to unambiguously communicate our findings, along with the mathematics that predicts them, to meet "basic requirements of science," as Bohr said in his reply to EPR [Bohr 1935, p. 697]. On the other hand, the ultimate nature of the reality responsible for observed phenomena may be and, in Bohr's view, is invisible to thought, and hence nothing about it can be communicated unambiguously or at all. The same situation is found in high-energy (QFT) regimes, whether we will ever be able to include the atomic constitution of measuring instruments in the theory or not. As Heisenberg says, following Bohr's argument, and aware of Bohr's exchanges with both Einstein and Schrödinger on the subject:

> Therefore, it has sometimes been suggested that one should depart from the classical concepts altogether and that a radical change in the concepts used for describing the experiments might possibly lead back to a nonstat[ist]ical [sic!], completely objective description of nature. . . . This suggestion, however, rests upon a misunderstanding. The concepts of classical physics are just a refinement of the concepts of daily life and are an essential part of the language which forms the basis of all natural science. Our actual situation in science is such that we do use the classical concepts for the description of the experiments, and it was the problem of quantum theory to find theoretical interpretations of the experiments on this basis. There is no use in discussing what could be done if we were other beings than we are. At this point we have to realize, as von Weizsäcker has put it, that "Nature is earlier than man, but man is earlier than natural science." The first part of the sentence justifies classical physics, with its ideal of complete objectivity. The second part tells us why we cannot escape the paradox of quantum theory, namely, the necessity of using classical concepts. [Heisenberg 1962, p. 56]

There is indeed no paradox here. Classical concepts reflect the essential workings of our biological and specifically neurological nature born with our evolutionary emergence as human animals. Our thinking, as the product of this machinery, is classical in that it is consistent with and leads to the concepts of classical physics. Any concept we form derive from and can only apply to observed phenomena, and quantum phenomena are physically classical as observed phenomena. They are different from classical phenomena because the data observed in them precludes us from describing how they come about by classical physics (which incapacity led to quantum theory) or in RWR interpretations, any physical theory or even making them available, visible, to thought. Such a conception, which would make the emergence of these data visible to thought, may be precluded by the same evolutionary biological or neurological nature of ours and, thus, by our classical thinking and language, developed in the interaction with (classical) objects consisting of millions of atoms, rather than anything on the atomic scale (e.g., [Heisenberg 1930, p. 11]). This is another manifestation (correlative to classical physics) of the fact that human nature and thus our thought are "earlier than natural science" and limit the latter. QM or QFT, however, allows one to probabilistically predict the data considered, without representing or even without us conceiving of how these data come about, or at least it allows for (RWR) interpretations, according to which QM or QFT does no more.

In classical physics we only need one theory for observing (or measuring), representing, and predicting the phenomena considered, which can be identified with the object considered, the interference of observation can be neglected. In both relativity and quantum theory (QM and QFT) we need classical theory to observe and measure the phenomena considered and the measuring instruments, but by relativistic and quantum theory, respectively, but with a crucial difference. In relativity we can, just in classical physics, still neglect the interference of measurement and, as a result, represent the behavior of the objects considered and predict this behavior, ideally deterministically. In quantum theory this interference cannot be neglected, essentially defining quantum phenomena as different from the objects considered, which makes our predictions, in general probabilistic, *regardless of interpretation*. In RWR interpretations, quantum objects or in the present view (in which quantum objects are only defined at the time of measurement by the Dirac postulate) the ultimate nature of reality responsible for quantum phenomena is placed beyond representation or conception. It is true that some classical theories, such as classical statistical mechanics or chaos theory, are probabilistic, but these theories are not fundamental because they do not deal with the ultimate constitution of nature. As explained, fundamental physics, as things stand now, requires three types of theories—classical, which do not consider the roles of both $c$ and $h$, relativistic (which are epistemologically classical), which do not consider the role of $h$ but do that of $c$, and quantum which must take into account $h$, and in high-energy regimes $c$, with both relativistic and quantum theories still using classical physics in representing the observable parts of measuring instruments and the outcomes of observations or measurements.

These considerations do not imply that new concepts, physical or (which is, however, not the issue at the moment) mathematical, are not possible in quantum theory. QM and QFT or their understanding and interpretation would not have been possible without the invention of new concepts, with Bohr's concepts of complementarity and phenomenon, or Schrödinger's concept of entanglement, among them. The question is whether one can avoid classical physical concepts or classical physics or whether new realist concepts, describing the ultimate nature of the reality responsible for quantum phenomena are possible or even necessary, as both Einstein and Schrödinger thought. On the first question, Bohr's or the present view is that classical concepts and classical physics cannot be avoided. On the second question, Bohr answers or at least that of the present author would be that such new realist concepts or theories *may not* be possible, which is not the same as *are not* possible. They might be possible. Complementarity and phenomenon are nonrealist concepts as concerns the ultimate constitution of the reality responsible for quantum phenomenon, but they contain realist components by involving the description of observed phenomena by ("old") classical concepts. Entanglement, defined as a concept mathematically, could, as concerns the physical reality considered, be understood along RWR lines as well, in accord with Bohr's view of the EPR experiment, manifesting entanglement and "suited to emphasize how far, in quantum theory, we are beyond the reach of pictorial visualization" [Bohr 1987, v. 2, p. 59]. Schrödinger himself spoke of the "entanglement of *predictions*," defined by the corresponding aspects of formalism, rather than quantum objects [Schrödinger 1935, p. 161; emphasis added].

Schrödinger was, however, not yet finished in his letter, and asked another question, which surprised Bohr as revealing something in Bohr's thinking, of which Bohr was not entirely aware himself at the time, and which is perhaps the most intriguing part of Schrödinger's letter:

> However that may be [as concerns a possible removal of the classical description of observation in relativistic quantum regimes], there must be clear and definite reasons which cause you repeatedly to declare that we *must* interpret observations in classical terms, according to their very nature. Whenever you say that, you state it so definitely and clearly, in the indicative, without any reservation like "probably", or "it might be", or "we must be prepared", as if this were the uttermost certainty in the world. It must be among your firmest convictions—and I cannot understand what it is based upon.
>
>   It could not be just the point (about which you talked so insistently to me already in 1926): that our traditional language and inherited concepts were completely unsuited to describe the phenomena with which we are concerned now. Because, in the course of the development of our science (and mathematics), from its earliest beginning to the situation at the end of the nineteenth century this was certainly the case over and over again. If the break with the old tradition seems greater now than ever before, then we should take into account that a particular time perspective is responsible for forming the impression *that* developments in which we ourselves take part, stands out as being more important and more essential that earlier ones, which we cite only from history, and whose stages we get to know mostly in reverse order. In fact, it is often difficult for us to imagine *earlier* ways of thinking. And although the difficulty of such a historical step *back* actually speaks most eloquently of *how* significant [the step] must have seemed to the pioneers of their earlier advances, still now and then we cannot avert feeling: "Incredible that, up to then, people were so narrow-minded!" Here, the underestimation of the time perspective shows itself most clearly.

Thus I think that the fact that we have not adapted our thinking and our means of expression to the new theory cannot possibly be the reason for the conviction that experiments must always be described in the classical manner, thus neglecting the essential characteristics of the new theory. (Letter to Bohr, October 13, 1935 [Bohr 1972-1996. v. 7, pp. 508-509])

Indeed, as Bohr's reply to Schrödinger, cited above, suggests, this is not the reason. It is not a matter of a going beyond a tradition, say as that of classical physics or even earlier quantum theory. It is difficult to object, and Bohr would not, to what Schrödinger says on this point. Hence, it would also be difficult to agree that Bohr was ever neglecting the essential characteristics of QM; quite the contrary, he affirmed them, not the least, as essentially different from classical physical theories, both deterministic or probabilistic. Bohr's emphasis of the classical description of measuring instruments is itself one of the essential characteristics of quantum theory, given that what is so classically observed can only be predicted by quantum theory. It is not that "we have not adapted our thinking and our means of expression to the new theory," because in fact physicists had so adapted their thinking (in terms of physical, mathematical, and even daily concepts), and Bohr was one of the first to do so. The reason for the conviction that "the experiments must also be described in a classical manner" are, as stated by Bohr in his reply: "*the seemingly obvious fact that the description of any measuring arrangement must, in an essential manner, involve the arrangement of the instruments in space and their functioning in time, if we shall be able to state anything at all about phenomena*." Or, more accurately, what is observed in experiments must be so described, because, as Bohr, added: "the argument here is of course first and foremost that in order to serve as measuring instruments, they cannot be included in the realm of application proper to quantum mechanics." This, too, then, is one the most essential features of quantum theory, which brings into our thought a relation to what is invisible to thought. But this relation would not be possible without describing in a classical manner what is visible to thought in quantum experiments.

Bohr's "*must*" in "we *must* interpret observations in classical terms" is stated "so definitively" and "without any reservation" because, while QM could become obsolete one day (although, remaining in place for a century now, not anytime soon in the author's view) or the RWR view, possibly in favor of realism, this necessity of interpreting observation in classical terms will remain. Schrödinger was astute to notice Bohr's "must," as Bohr didn't fail to acknowledge this point in his reply: "I found it most amusing that you noticed—which I myself had not at all been aware of—that just on this point, and only on this one, I do not say, 'it might be'" (Letter to Bohr, October 13, 1935 [Bohr 1972-1996. v. 7, p. 512]). Bohr will be more aware of this fact from this point on, with its significance even more pronounced in his subsequent writings. Bohr's "must be" reflects his assumption of the necessity of unambiguous, and in this sense objective, communication of, along with the logical and mathematical structure of quantum theory, the outcomes of experiments, insured by the classical description of the observable part of measuring instruments, in accordance with the Bohr postulate. As he said later (in 1949):

It is decisive to recognize that, *however far the phenomena transcend the scope of classical physical explanations, the account of all evidence must be expressed in classical terms*. The argument is simply that by the word "experiment" we refer to a situation where we can tell others what we have done and we have learned and that, therefore, the account of the experimental arrangement and the results of the observation must be expressed in unambiguous language with suitable application of the terminology of classical physics. [Bohr 1987, v. 2, p. 39][20]

Our expectations or probability assignments concerning such outcomes may be different, depending on different information we have pertaining to a given experiment, and in this sense, they are subjective or personal. The latter might be a better concept insofar as these assignments are shaped by things in the world, such as measuring instruments or the world itself which is assumed in this article or by Bohr to exist independently and thus to be external to an agent. Things are rarely, if ever, completely subjective, permitting that such exterior factors are interiorized at the time of an assignment of one or another probability to a future event. There is nothing paradoxical or inconsistent with Bohr's claims in this understanding of probability. In life, too, we can have different expectations concerning future events given the information we possess (which may be different), although QM, as a mathematical-experimental science, gives us a precise probability calculus to predict quantum events, which is, again, all it does in the present view. Life rarely gives us such means.

    At the same time (hence the consistency with Bohr's claims), any measurement, in any quantum experiment that would be performed would give a definitive, visible and informationally communicable outcome, and as such is classical. An agent cannot control it but can only predict it probabilistically by means of QM (cum Born's rule). One might be able to decide (although it may not be simply a matter of a free choice by our consciousness or even unconscious) which observation or measurement to perform, for example, one or the other complementary observation, but one cannot control the specific outcome of it as concerns which value one obtains, even if one controls the preparation of the instrument that will register that outcome. In addition, as noted, one can always perform an alternative, complementary, measurement at the end point of the experiment, which will irrevocably disable the original estimate. (In classical mechanics, one can, again, always measure and predict, deterministically, all variables necessary for accounting, representationally, for the system considered.) This makes measurement objective in this double sense—the lack of control of an outcome and the possibility of an unambiguous communication of an outcome—but in the present view, only in this double sense, rather than *objectively* attributing anything to nature itself, apart from its existence and, as part of it, human existence. Making an observation or measurement is, as stated, a unique act or event of creation with a unique outcome that can be performed by a particular agent or several agents and as such has subjective or, again, personal aspects, including those shaping our decision concerning this action, a decision inherent in the very idea of experiment [Bohr 1935, p. 699]. Once the measurement is performed, however, the outcome becomes fixed as a permanent record, part of the archive of physical data, always classical and visible to thought or even our immediate phenomenal perception. It may be unknown to others, but that it is not the same as being or bound to remain subjective. It is true, too, that, as any record, it must still be experienced as such by us or others to be meaningful. Thus, performing an *act* of observation or measurement is personal (if sometimes determined collectively), but its *outcome* need not be. It can also be experienced differently by different agents, and in this sense, it is always personal and, in the first place, human.

    Science is a human enterprise. But sharing and communicating our estimates of possible events and experiences is also human and doing so is helpful and even unavoidable in human life. Science capitalizes

---

[20] As indicated above (note 8), it is possible to argue that, while necessary for the description of the observable quantum phenomena and measurements associated with them, classical physics is not a separate theory but rather a limit case of QM, thus eliminating it from the class of fundamental theory, which is, however, not the present view or, I would argue, that of Bohr. Thus, in the present view, if considered by itself, the cat in the cat experiment, is always a classical object that cannot be handled, either representationally or (which is only possibility in RWR interpretations) predictively, by QM. QM is only applicable in the cat experiment because there is a properly quantum aspect to it: the emission of a particle by the radioactive atomic substance used. As will be seen, the same considerations apply in the Wigner's friend experiment, sometimes used to argue that everything can be considered as quantum, without any use of classical physics. Most of these arguments, moreover, contend or imply that the observable parts of measuring instruments can also be handled by QM, without, in contrast to Schrödinger's (subtler) argument that the situation requires new concept beyond both classical and quantum physics.

on this fact and on the possibility that the communication involved may be made unambiguous, helped by the use of mathematical symbols, central to modern physics, from Galileo on. These symbols, too, or their organization are visible to thought and hence unambiguously communicable, including those of the mathematical formalism of QM or QFT. Mathematics itself, as a discipline, depends on this fact. In classical physics and relativity, however, how the outcomes of experiments come about is visible to thought as well, and may be assumed to be independent of observation, for all practical purposes, but, in the present view, still only for all practical purposes, defined by human agents and agencies, such as science. Not so in quantum physics, essentially dealing with and fundamentally shaped by what is invisible to thought. In quantum physics, the role of human agents and experimental technology cannot in principle be neglected, as reflected in the nature of quantum probability, which, as discussed above, is no longer due, as in classical physics when it uses probability, to our *insufficient* knowledge of how the phenomena considered come about. At stake in RWR interpretations is the impossibility in principle of any knowledge or even conception concerning how this happens, which makes probability fundamentally irreducible. The mathematics of quantum mechanics is visible to thought, and as such is unambiguously communicable. But how what this mathematics predicts (in general probabilistically) comes about, as outcomes of quantum experiments, is not.

We do not know what Schrödinger thought upon receiving Bohr's reply, although it appears that he had never have accepted Bohr's view concerning the irreducible role of classical concepts in quantum theory. It is not clear, for example, to what degree, if any, the cat paradox or even his paper overall were an attempt to show that new physical concepts may after all be necessary in quantum theory. As explained below, it appears that, by saying that "the $\psi$-function of the entire system would express this by having in it the living and the dead cat (pardon the expression) mixed or smeared out in equal parts," he assumes the cat to be a quantum object [Schrödinger 1938, p. 157]. This view has a difficulty in the fact that, if we open the box or use a box with glass walls, we will see the cat, the *same* cat, at any stage of the experiment, while one can never so see a properly quantum object, such as an electron, for detecting which one always needs an instrument. In the present view, moreover, a quantum object is a concept that only applies at the time of the experiment by the Dirac postulate, and hence implies that each observation concerns a different quantum object, although identifying these objects is permissible in low-energy (QM) regimes, but not high-energy (QFT) regimens. In the cat experiment, we always see the same cat, which can, as noted, change its state, but not its sameness, always visible to thought. Be it as it may on that score, Schrödinger thought, as did Einstein, that new concepts associated with quantum objects and their behavior might be necessary received a new support from the EPR experiment. Such concepts, they thought, would ground a realist alternative to QM, viewed by Schrödinger as "perhaps after all a convenient calculational trick" [Schrödinger 1935, p. 167]. It is difficult to assume that Einstein saw it as anything more than that. Neither thought that QM was likely to be interpreted on realist lines, although such interpretations have been advanced. For Bohr, as explained, the EPR experiment confirmed, in accordance with his (strong RWR) interpretation, "how far, in quantum theory, we are beyond the reach of pictorial visualization," to the point of reaching what is invisible to thought [Bohr 1987, v. 2, p. 59].

**4. Schrödinger's cat experiment through the optics of visible and invisible to thought**

Schrödinger's paper containing the cat experiment was a response to EPR's paper, which it discusses at some length, and was, arguably, most important for the concept of entanglement, introduced by Schrödinger, and its overall discussion of QM, its main concern, as reflected in its title "The present situation in quantum mechanics" [Schrödinger 1935]. His analysis is thoroughgoing and penetrating, even though (or perhaps because( Schrödinger assessed QM, especially as interpreted along the (Copenhagen) RWR lines, as "a doctrine born of distress" [Schrödinger 1935, p. 154]. He saw QM, if not necessarily as incomplete insofar as concerns its capacity to predict all that was possible to predict (or else nonlocal), as EPR argued, but then as "perhaps after all only a convenient calculational trick" [Schrödinger 1935, p. 167]. EPR's experiment gave Schrödinger, as it did to Einstein, new hopes that an alternative realist theory of quantum phenomena might be possible. The cat experiment was part of Schrödinger's overall analysis of QM, a relatively marginal part, which did not appear to have initially received much attention. Neither did initially the paper itself, even the concept of entanglement, introduced there, a major contribution to QM. During the last half a century or so, however, the cat experiment has been interminably discussed in technical, philosophical, and popular literature, and has even acquired a semi-mythical status. There are many reasons for this upsurge of attention to it, such as its role in helping realist or classical causal views of QM or fundamental physics, or countering the Bohr postulate, often resisted as much as the lack of realism or classical causality (e.g., [Plotnitsky 2022b]). There is of course also a narrative appeal to the experiment, especially in popular accounts, but not only there.

From the present perspective, there is nothing especially remarkable or revealing in the cat experiment, or anything that would challenge Bohr's or the present interpretation. There does not appear to be any evidence that Bohr ever commented on the experiment or on Schrödinger's paper. The letter exchange, discussed above, between Schrödinger and Bohr concerning Bohr's emphasis on the classical description of measuring instruments is relevant to the cat experiment. But, as preceding discussion makes clear, this exchange took place before Schrödinger's paper was published and was about Bohr's views, rather than any aspect of Schrödinger's paper. I'd surmise that Bohr would not find anything in the paradox either of much interest or as challenging his views. I'd also surmise that he was likely to have seen, as I do here, the cat as a classical and not a quantum object. As indicated above, while not assumed by Bohr, the Dirac postulate, which only applies to quantum and not to classical objects, lends further support to this view. This is because the postulate states that each quantum observation concerns a different quantum object, while only allowing one to assume that successive observations deal with the same quantum objects as a statistically permissible idealization of low energy (QM) regimes but not in high-energy (QFT) regimes. By contrast, the cat is aways the same object (if in a different classical state) at any stage of the experiment. At least, as I shall argue, it is difficult to assume otherwise. According to Schrödinger, then:

> One can even set up quite ridiculous cases. A cat is penned up in a steel chamber, along with the following diabolical device (which must be secured against direct interference by the cat): in a Geiger counter there is a tiny bit of radioactive substance, so small, that perhaps in the course of one hour one of the atoms decays, but also, with equal probability, perhaps none; if it happens, the counter tube discharges and through a relay releases a hammer which shatters a small flask of hydrocyanic acid. If one has left this entire system to itself for an hour, one would say that the cat still lives if meanwhile no atom has decayed. The first atomic decay would have poisoned it. The $\psi$-function of the entire system would express this by having in it the living and the dead cat (pardon the expression) mixed or smeared out in equal parts. [Schrödinger 1935, p. 157]

In the present interpretation the last sentence would not apply, at least as stated, and the preceding two sentences, which are in accord with the present view, appear to contradict the last sentence. I shall explain why this is so presently. First, however, Schrödinger adds an elaboration that is rarely discussed or given proper attention, which provided a further context for his thought experiment. He says: "It is typical of these cases that an indeterminacy originally restricted to the atomic domain becomes transformed into macroscopic indeterminacy, which can then be resolved by direct observation. That prevents us from so naively accepting as valid a 'blurred model' for representing reality" [Schrödinger 1935, p. 157]. A blurred model is defined by a view of the $\psi$-function as "an imagined entity that images the blurring of all variables at every moment [unless a measurement intervenes] just as clearly and faithfully as the classical model [images] its sharp numerical values" [Schrödinger 1935, p. 156]. In other words, the problem arises if one sees the $\psi$-function as representing the independent behavior of quantum systems, in this case as blurred. In the present view, the $\psi$-function does not "faithfully" represent the behavior of the quantum object considered or the ultimate reality responsible for quantum phenomena, because it does not represent this reality at all. It only provides an (discrete) "expectation-catalog" for possible future experiments, as Schrödinger himself called it [Schrödinger 1935, p. 154]. In developing his wave mechanics, Schrödinger initially aimed for a (wave-like) representation of the ultimate reality responsible for quantum phenomena in his project for his wave mechanics that led him to his famous equation. He had, however, long given up on the idea by this point in view of the difficulties of reconciling his wave mechanics had with observable features of quantum phenomena, in particular their discreteness and the probabilistic nature of predictions concerning them. His equation itself can of course be and has been (immediately in the Göttingen-Copenhagen circles) interpreted so as to accommodate these features, especially given M. Born's probabilistic interpretation of the $\psi$-function, eventually part of RWR interpretations of QM.

These interpretations, including the present one, would, however, contrary to Schrödinger's statement, preclude the claim that "the $\psi$-function of the entire system would express [the situation] by having in it the living and the dead cat (pardon the expression) mixed or smeared out in equal parts," unless Schrödinger meant that his claim only applies to blurred variables. His claim, however, appears to be more general and applicable when one does not view the variable considered as blurred, as his subsequent reference to the cat experiment indicates [Schrödinger 1935, p. 161]. In any event, in the present view, the cat, as a classical object, would always be either dead or alive at any stage of the experiment, as Schrödinger's preceding sentences imply: "If one has left this entire system to itself for an hour, one would say that the cat still lives if meanwhile no atom has decayed. The first atomic decay would have poisoned it." Why then claim: "the $\psi$-function of the entire system would express this by having in it the living and the dead cat (pardon the expression) mixed or smeared out in equal parts"? It is also, in principle, possible that by his phrasing "the $\psi$-function of the entire system *would express this by having in it*" (emphasis added) Schrödinger only meant the mixing of amplitudes for these two outcomes so that "the $\psi$-function of the entire system" contains both possible future outcomes, as would be the case in the present view. Schrödinger, however, does not qualify his statement in this way. In the present view, without any conflict with the first two sentences, which only refer to a classical object, what will be "mixed" or superposed are mathematical state vectors in the formalism. This mixture enables the probabilities of predicting the atomic decay involved, to which, as a quantum process, such terms as "dead" or "alive," or any other terms, do not apply. It is only because of this purely mathematical mixture that one is able to estimate the probability of finding the cat dead or alive. The $\psi$-function has no association with the cat apart from these predictions (via Born's rule), and it never represents the state of the cat, as a classical object. The $\psi$-function never represents the physical state of a quantum object either, as would be implied, by suggesting that the cat is seen as a quantum object, by Schrödinger's formulation.

In the present view, the cat is never mixed or smeared in equal parts between the living and the dead cat. It is either the alive or dead cat at any stage of the experiment. One merely does not know (after a certain moment in time, while cat is inside the box) whether it is alive or dead until one opens the box. There is nothing that can be said or thought of concerning the *ultimate* reality responsible for quantum phenomena (including, quantum objects, in the present view defined only at the time of observation by the Dirac postulate), including that responsible for the atomic decay in the cat experiment. By contrast, there are always things we can say about any classical reality, involved in quantum experiment, as part of what is, *in principle observable*, in them, as is the cat in the cat experiment.[21] The former reality is invisible to thought, the latter is visible to thought.

---

[21] The emphasized phrase deliberately echoes Heisenberg's famous and much misunderstood, especially along empiricist (Machian-like) lines, opening claim in his first paper of QM to the effect that he aims to ground his new mechanics in "the *relationships* between quantities which in principle are observable" [Heisenberg 1925, p. 263]. The quantities in question are empirically observable in measuring instruments, but the *relationships* in question (the word usually disregarded in empiricist readings of this statement) are the probabilistic relationships established by his new mechanics. As Heisenberg said, shortly before completing his paper: "What I really like in this scheme is that one can really reduce *all interactions* between atoms and the external world ... to transition probabilities" [Heisenberg, Letter to Kronig, 5 June 1925; cited in Mehra and Rechenberg 2001, v. 2, p. 242]. By speaking of the "interactions between atoms and the external world," this statement suggests that QM was only predicting the effects of these interactions observed in measuring instruments, without representing how these effects come about. As explained, this procedure replaced measurement in the classical sense (of measuring some preexisting properties of

One could, however, in the arrangement of the cat experiment, consider the cat as *part* of an object under investigation, concerning which the prediction in question is made by means of QM, but as explained earlier, only as part of this object because a proper quantum object must be involved in order to have a quantum experiment. The cat experiment is a quantum experiment because of the radioactive decay and not because of the cat. Therefore, considering the cat as part of an object under investigation does not change the point that the cat is a classical object, always visible to thought or to our immediate sense perception (before and after the experiment, or throughout if, if the box has glass walls), but not a quantum object, which is never available to our phenomenal perception and, in the present view, is invisible to thought. One can at any point see the cat as such, independently of a quantum observational device, by opening the box or, again, using the box with glass walls, but one can never see a quantum object as such or rather (since it cannot be seen) establish its presence without a suitable observational device. A quantum object cannot be observed as separated from the phenomenon considered, which is a result of the interaction between this object and the instrument. It would, accordingly, be more reasonable to see the cat as a classical object, while, within in the overall arrangement of the experiment, being part of the object of investigation by quantum means, enabling one to predict its possible classical state of being dead or alive at the final stage of experiment.

As such the cat is also an object that can be described by ordinary language, as opposed to a quantum object like an electron, which is, in RWR interpretations, merely a name, without a concept attached to it. This fact makes misleading using, as is done sometimes, such notations as $|\psi\rangle = \alpha|dead\rangle + \beta|alive\rangle$, as opposed to the something like $|\psi\rangle = \alpha|h\rangle + \beta|v\rangle)$. The later refers to state vectors, in a superposition, used to predict definitive classical events (which are never in a superposition), such as, within the chain of events in the arrangement, that of the cat being dead or alive in the cat experiment, but has no other connections to either (physical) state of the cat. Technically, QM predicts the effects that quantum objects (or in the present view, the ultimate reality responsible for quantum phenomena) can have on the classical world we experience. These effects define quantum phenomena or events. As any observation in quantum physics, opening the box in the cat experiment is the phenomenon that reveals a classical state of the reality, a state in this case already established in advance, which includes the cat, either dead or alive. One or another properly quantum object, such as a radioactive atom and a particle it emits is always necessary to have such an effect, even if the object under investigation, as different from a measuring instrument, can contain a classical object, such as the cat. Calculating the probability of any such prediction will have to involve *h* because of the radioactive decay as involving properly quantum objects, while the cat is of no help in estimating such probabilities.

To support the case just outlined more rigorously, I turn to Bohr's argument in his reply to EPR, concerning "discriminating in each experimental arrangement between those parts of the physical system considered which are to be treated as measuring instruments and those which constitute the objects under investigation," and the question of the cut thus arising. According to Bohr:

> This necessity of discriminating in each experimental arrangement between those parts of the physical system considered which are to be treated as measuring instruments and those which constitute the objects under investigation may indeed be said to form a *principal distinction between classical and quantum-mechanical description of physical phenomena*. It is true that the place within each measuring procedure where this discrimination is made is in both cases largely a matter of convenience. While, however, in classical physics the distinction between object and measuring agencies does not entail any difference in the character of the description of the phenomena concerned, its fundamental importance in quantum theory … has its root *in the indispensable use of classical concepts in the interpretation of all proper measurements*, even though the classical theories do not suffice in accounting for the new types of regularities with which we are concerned in atomic physics. In accordance with this situation there can be no question of any unambiguous interpretation of the symbols of quantum mechanics other than that embodied in the well-known rules which allow us to predict

---

quantum objects) with establishing, by using measuring instruments, quantum phenomena, which can be treated classically without measuring the properties of quantum objects, a view was adopted and developed by Bohr.

the results to be obtained by a given experimental arrangement described in a totally classical way. [Bohr 1935, p. 701; second emphasis added]

It is important to avoid two common misunderstandings of this and related statements by Bohr. The first concerns measuring instruments, in view of Bohr's insistence on the classical description of the observable part of measuring instruments, a subject discussed in detail in Sections 2 and 3, beginning with the fact that instruments have quantum parts through which they interact with quantum objects. The second concerns quantum objects. Bohr's statement does not mean that, while observable parts of measuring instruments are described by classical physics, the independent behavior of quantum objects is described by means of the quantum-mechanical formalism, which assumption would be in conflict with the RWR interpretation held by Bohr. This type of (realist) view has been adopted by some, sometimes under the heading of "the Copenhagen interpretation," beginning, influentially, with Dirac's and von Neumann's classic studies, with Dirac's book originally published in 1930 and von Neumann's (in German, in 1932 [Dirac 1958, von Neumann 1955]. Both books, moreover, assume a classically causal independent behavior of quantum objects, with probability brought in only by measurement. This was, however, not Bohr's view, especially at this stage of his thinking in 1935, or even almost immediately after the Como lecture of 1927, which may be seen as having adopted, still ambivalently, this type of view and which arguably influenced both Dirac and von Neumann in this regard [Plotnitsky 2016, pp. 198-211]. In the passage in question, Bohr only says that classical theories cannot account for how quantum phenomena (physically described classically) come about or predict what is observed. He does not say that the independent behavior of quantum objects or objects under investigation (which may not be quantum but must contain quantum objects) is represented by the formalism of QM. In Bohr's view, the "symbols" of QM only have a probabilistically predictive role, without, by the Heisenberg postulate, offering a representation of how quantum phenomena come about, while quantum phenomena themselves are represented by classical physics. So, QM does not represent them either, by the Bohr postulate. Thus, in Bohr's interpretation, while predicting, in general probabilistically, the data observed as part of phenomena, the formalism of QM is otherwise dissociated in physical terms from both the ultimate nature of reality responsible for quantum phenomena and these phenomena themselves, which phenomena are described by classical physics.

The circumstance that "the place within each measuring procedure where this discrimination is made is … largely a matter of convenience" is related to, although is not quite the same as, the arbitrariness of the cut or the Heisenberg cut, or sometimes the Heisenberg-von-Neumann cut, because Heisenberg and von Neumann favored the term (each giving it a somewhat different meaning), not used as such by Bohr. Bohr qualifies this claim, and this qualification is important, including in the context of the cat experiment. While "it is true that the place within each measuring procedure where this discrimination is made is … largely a matter of convenience," it is true only largely but not completely, because "in each experimental arrangement and measuring procedure we have only a free choice of this place within a region where the quantum-mechanical description of the process concerned is effectively equivalent with the classical description" [Bohr 1935, p. 701].[22] Thus, the ultimate constitution of the physical reality or quantum objects and in quantum part of the instruments interacting with quantum objects is never on the measurement side of the event, and by the same token they can never serve as measuring instruments either. As beyond representation or even conception, as invisible to thought, quantum objects cannot be assigned any properties, even at the time of measurement. This impossibility is correlative to their position of always being on the other (than measurement) side of the event. All observable properties, are, by the Bohr postulate, only those of the observable parts of measuring instruments, described classically, but appearing under the impact of quantum objects. QM only predicts these visible properties. In Bohr's or the present view, in part by virtue of associating the cut with the discrimination between what is considered the object under investigation and what is considered as the measuring instruments, the cut and its (within certain limits) shifting nature need not imply that the classical theory, say, classical mechanics or classical electromagnetic theory is a special form of QM or (in the case of electromagnetism) QFT. As emphasized throughout this article, in Bohr or the present view, or that of Heisenberg, these are two different theories that deal with two different types of objects, even though classical objects are still composed of quantum objects or, in the present view (because the concept of a quantum object only applies at the time of observation) of the same ultimate reality. We do not know how classical objects (which could be considered, at least in principle, independently of observational technology) emerge from this this ultimate reality. It is true that a quantum system with a large number of coherent (quantum) states behaves close to a classical system, close but a) not quite; and b) still it is only a special class of quantum systems, which are still not the same as classical systems, which one uses to describe classical objects, including measuring instruments.

---

[22] This situation may be seen as a manifestation of Bohr's correspondence principle, according to which the quantum-mechanical and the classical descriptions give the same predictions in the classical limit.

As indicated earlier, the cut, as here understood, reflects the possibility of placing some classical parts of the overall arrangement in a quantum experiment (including the cat experiment) on either side of the cut, as "an object under investigation," concerning which predictions can be made by QM. As I argue, however, the arrangement must include a quantum object (like a particle in radioactive decay in the cat experiment) for the experiment to be quantum, to have quantum effects. Thus, the cat is always visible at least to our mind's eye, even when, while inside the box, not actually available to our sense perception. A quantum object is never available to such a perception, and one always needs an experimental device (which could be multi-stage, as in the cat experiment) capable of interacting with this object to have a quantum effect. This effect is manifested in classically observed phenomenon or event, such as the cat being alive or dead, in this case in fact, rather than being a properly quantum effects, preceded and made possible by another classically observed event, the breaking of the flask, which is more properly quantum effect, due to the interaction between it and the emitted particle. This interaction is quantum, while the rest is the chain of classical events triggered by it. Bohr, accordingly, does not call a composite object, containing both classical and quantum objects, on the object side of the cut a "quantum object," but the "object under investigation." Quantum objects, while they can also be objects under investigation, are only those objects that strictly belongs to the ultimate reality responsible for quantum phenomena, and they are always on the object, never measurement, side of the event, and hence a quantum object can never be a measuring instrument either. Other parts of objects under investigation in quantum experiments are physically classical objects, such as the cat or everything inside the box, except that part of the flask that can interact with the emitted particle. While it can be *part* of an object of investigation in a quantum experiment and treated by quantum means, if considered by itself, a classical object cannot be treated as a quantum object. Any prediction or measurement associated with a quantum object, elemental (such as an electron) or composite (such as a Josephson device), will always involve $h$, thus correlative to what is invisible to thought in quantum physics, even though $h$ itself is observed classically. Observing the cat in the cat experiment does not require $h$ because the cat is a classical object, which, again, cannot be treated as a quantum object, only a quantum part of it can, like protons in its body in the MRI test.

On the other hand, in certain circumstances, a quantum object could be *treated* for all practical as a classical object, *but without ever being a classical object*. Thus, as when it is far enough from the nucleus (for large quantum numbers), an electron can be treated as behaving classically. This is, however, an approximation or idealization which disregards possible quantum effects of this behavior. As Bohr noted, also connecting this situation to "mechanical pictures" and "classical pictures," as visible to thought:

> [I]n the limit of larger quantum numbers where the relative difference between adjacent stational states vanishes asymptotically, mechanical pictures of electronic motion [as orbits] may be rationally utilized [by the correspondence principle]. It must be emphasized, however, that this connection cannot be regarded as a gradual transition toward classical theory in the sense that the quantum postulate [as an essential discontinuity and individuality of quantum phenomena] would lose its significance for high quantum numbers. On the contrary, the conclusions obtained from the correspondence principle with the aid of classical pictures depends just upon the assumptions of the conception of stationary and of [discrete] individual transition processes are maintained even in this limit. [Bohr 1987, v. 1, p. 85]

By this point (in 1927), Bohr adopts the view, which, following Heisenberg, equally renounced both the classical, orbital "picture" of stationary states, still assumed in Bohr's 1913 theory, and any classical view of the transitions, "quantum jumps," between states, already abandoned by Bohr's 1913 theory, the first instance of the RWR view (even if only partially applied). The concept, while it enabled Bohr to account for the stability of atoms (vs. E. Rutherford's preceding view), was nevertheless incompatible with classical mechanics and classical electrodynamics alike. Neither the time nor direction of each jump could be explained, although it could be predicted probabilistically or statistically, which, thus, from a classical perspective paradoxically, brought the atomic stability and quantum randomness together. This stability is of course that of a dynamic system, which can change its states, although these changes could only be registered in measuring instruments. Bohr's conceptual framework makes the term "jump" misleading in suggesting some representation of what happens. Electrons do not jump; quantum states (as physical states) discontinuously change, and no representation of how they do this is available. What was responsible for these changes was assumed to be real, but this reality was assumed to be at least beyond representation, in accord with the weak RWR view, although intimating the strong RWR view, insofar as no concept of how these transitions appeared to be possible to form either. In Heisenberg's approach leading him to his invention of QM, the same situation defined the case of electrons in stationary states. Electrons were not moving in orbits around nuclei: their quantum states (associated with variables other than energy, as the energy remained the same in a stationary state) were changing, with these changes observable as discrete phenomena. In this view, there were only the states of quantum objects, manifested in measuring instruments, and transitions between these states. This was a decisive shift in our understanding of the nature of physical reality.[23] One might say that, rather than making any transition to a new energy, an electron in a given stationary state disappears and a new electron is born in this new stationary state. Each corresponding measurement will detect a different electron, in accord with the Dirac postulate. The wave function for an electron in an atom can be recast in terms of annihilation and creation operators, used in QFT.

Bohr's statement cited above concerning the behavior of electrons in the case of large quantum numbers confirms his view of classical objects and processes as visible to thought or even our immediate phenomenal perception, and the behavior of quantum objects as, at this point (in 1927), no longer at least to our general phenomenal intuition, and if not yet invisible to thought, as Bohr came to understand the situation by the late 1930s. The behavior in the limit of large quantum numbers can be *treated* for all practical purposes as that of classical objects. This treatment is, however, merely a workable approximation of what is the ultimate nature of the reality responsible for what is thus observed, a reality invisible to thought, and one still require a measuring instrument for this observation. At bottom one still deals with the combination of stationary states and discontinuous quantum jumps. These states are too close to each other for this combination to be detected, but one would, as it were, register these states (as invisible to thought electrons or photons they emit still cannot be "seen") by "zooming" on them, if one had an instrument to do so. Any such instrument would, however, need to be able, by interacting with electrons or "emitted" photons, to register properly quantum effects. Technically, an "emission," too, is a classical concept, which cannot represent how photons are "emitted," which is invisible to thought. All we can see are traces of photons, or what we assume to be photons, traces manifested, literally visible, in spectra. Similarly, a macro quantum object (still defined as such by its microscopic quantum constitution), such as a Josephson device, can only be detected as quantum by means of a suitable instrument. Otherwise, it will be observed as a classical object and as such as something (two superconductors standing in a lab) available our immediate phenomenal perception.

Thus, in quantum physics, on the one hand, there is always a discrimination between an object and an instrument, and, on the other, their indivisibility in quantum phenomena, or what Bohr calls the wholeness of phenomena or its closed nature, from which one can never extract the object itself at the time of measurement. Any investigation in quantum theory must involve this combination, which is also that of what is invisible to thought and as such cannot be communicated unambiguously, and what is visible to thought, via observational instruments, and can be communicated unambiguously. This situation thus sharply contrasts with that of classical physics or relativity, where the role of measuring instruments can be neglected or controlled and where, as a result, which always deal with what is visible to thought and, as such unambiguously communicable or sharable as information. If the object under investigation is classical (visible, representable, with its character unambiguously communicable, and so forth), like the cat in the cat experiment, it can always be considered independently apart from quantum experiments and discussed unambiguously. There is, as noted, never any ambiguity in assessing the cat as an independent object in the cat experiment but only two unambiguously defined possibilities, each visible to our mind's eye, of the cat being either dead or alive, with the probability defined by the $\psi$-function associated with the atomic decay and only secondarily to the state, always classical, of the cat. A cat, inside or outside the box, is always a cat, dead or alive. As such it can only be seen as a physically classical part of the arrangement, inside the box, before the interaction with the particle emitted by an atom, which can never be observed (and terms like particle or emission cannot apply in any sense we can attribute to these terms). The cat can be on both sides of the event (or the cut), but the radioactive decay or the particle emitted by it can only be on one side, the side of the object, and never the measurement side. This emission occurs or not regardless of the cat in the box, or the box, or the flask, all of which are classical and are parts of the arrangement made by us, while the radioactive atom is prepared by nature. The cat could be removed from the box in advance without affecting this possible quantum event. The flask is the only classical object that interacts with the emitted particle, which enable one to register with the presence of the emission, if it occurs.

---

[23] I am indebted to Laurent Friedel on this point.

Technically, one need not see the opening the box as a quantum experiment, as the properly quantum experiment in the arrangement, which is the shattering of the flask, occurs (if it does) before the box is opened, and then the outcome of this experiment leads to the event that classically affects the cat. The cat is more like an "agent," akin to (although of course not the same as) "Wigner's friend" in a related famous experiment, than an instrument, and is, again, never a properly quantum object.[24] Neither, again, is anything else in the arrangement, except the flask, inside the box. But, as explained, even if one does see the whole arrangement and an instrument (an arrangement and an instrument made by us as humans), the cat, just as the box, it is still only a classical part of this arrangement, always visible to our mind's eye or directly if the box has glass walls, even if the arrangement requires us to use QM to predict what happens once one opens the box. The cat or the friend in the Wigner's friend experiments is not an instrument either, and either arrangement implies the presence of an observational instrument capable of interacting with quantum objects and registering (classically) the outcome of such interactions. The cat does not, of course, consciously observe such an outcome in the way a human agent would. The cat can only manifest this outcome by being dead or alive. One the other hand, the friend in Wigner's friend experiment does consciously observes it, which fact was central to Wigner's original argument. This difference, however, does not change the fact that the cat or the friend is a classical object or that both experiments essentially depend on the role of properly quantum objects, which are never of the measurement side of the event. Nor could they serve as measuring instruments, which requires to have a classical part in order to be observed by an agent. Nor of course could quantum objects serve as agents.

---

[24] Although it has additional complexities, the Wigner's friend experiment [Wigner 1961] can be considered along the lines of the argument advanced here. In Wigner's scenario, "the friend" hidden from "Wigner" inside some lab (just as the cat is hidden from the observer in the cat experiment), performs an experiment on a previously prepared quantum system, $S$, with the outcome, which, unlike the initial preparation, is hidden from "Wigner" as well. "Wigner" leaves the lab after the initial preparation, which enables one to associate with $S$ (which is, in the present view, not the same as assign to $S$) the wave function $|\psi\rangle$, known to both. QM can, then, be used by "Wigner" in estimating this hidden outcome. This is, I would argue, possible while, just as in the present view of the cat experiment, considering "the friend" as a classical object within the overall arrangement, which, as that of the cat experiment, contains a proper quantum object, $S$. The case would require a separate analysis. I might note, however, that most discussions of the Wigner's friend experiment and the problems and paradoxes found in many of them, beginning with Wigner's own encounter assume that "the friend" (or sometimes "Wigner") can be considered as a quantum object. For more recent treatments, see [Pusey 2018, Bauman and Brukner 2020, DeBrota et al 2020], and further references in these articles. It is not my aim to assess these arguments (sometimes questioning each other) as concerns their effectiveness in resolving the "paradoxes" of Wigner's experiment and Wigner's own initial argument, which, I would argue, in effect suggests, even if against Wigner's own grain, the difficulty of assuming that the friend is a quantum object. Another, related, feature of some of recent arguments, most especially [DeBrota et al 2020], which is based in quantum Bayesianism (QBism), is their claim of the subjective nature not only of our predictions, a view assumed here as well, but also of the outcomes of quantum measurements, a view not assumed here. In the present view, following Bohr, these outcomes are objective in the sense of being unambiguously communicable (with further qualifications given above). This assumption is correlative to that of the physically classical description of the observable part of measuring instruments and quantum phenomena. In the present view, "the friend" is a classical object, just as is the cat in the cat experiment, even though the arrangement considered can be treated by QM as concerns "Wigner's" estimates of the outcome of the friend's measurement.

## 5. Conclusion

I return, in closing, to Denmark, first, not to that of Bohr but that of Shakespeare and *Hamlet* three centuries earlier, and the lines of the play, used as my first epigraph:

> Hamlet [commenting on his dead father]:
> My father—methinks I see my father.
> Horatio:
> Where, my lord?
> Hamlet:
> In my mind's eye, Horatio.
> [*The Tragedy of Hamlet*, *Prince of Denmark,* Act 1, Scene 2, ll. 183-185]

The reason for Horatio's puzzlement is that he saw the ghost of Hamlet's father and wondered if perhaps Hamlet has already seen the ghost as well, which Hamlet's response proves not to be the case. Hamlet's encounter with the ghost of his father is yet to come. At stake in this scene is Hamlet's image of his father in his mind's eye, and thus as something visible to thought. This image shadows Hamlet and the play from beginning to end, and Hamlet's encounter with the ghost, dramatic and consequential for the play, adds to the power as this image in and over Hamlet's mind, but this power was already there all along. I am, however, not concerned here with much discussed psychological, such as psychoanalytic, implications of this power, but instead with the capacity of our thought, conscious and unconscious, to create an image of the world and of objects in the world, on which Shakespeare capitalizes in *Hamlet* and his other works. No less remarkable, however, is our thought's capacity to think that which is beyond thought, is invisible to thought, and hence has no image in our mind's eye, such as the ultimate nature of the reality responsible for quantum phenomena. Shakespeare might have realized this capacity of thought at least to some degree, as suggested by Hamlet's comment to Horatio after his encounter with the ghost:

> Horatio: O day and night, but this is wondrous strange!
> Hamlet: And therefore as a stranger give it welcome.
> There are more things in heaven and earth, Horatio,
> Than are dreamt of in your philosophy.
>     [*The Tragedy of Hamlet, Prince of Denmark,* Act I, Scene 4, ll. 165-166]

Some editions have "our philosophy." "Your philosophy" makes Hamlet more suspicious of philosophy. It makes him more akin to a quantum physicist, who can only estimate the probabilities of future events defined by experiments Hamlet stages at the castle of Elsinore, a prominent aspect of the play. There are, quantum physics may indeed be telling us, things in, or beyond, heaven and earth that we cannot dream of or otherwise see in our mind's eye, consciously or unconsciously.

    Bohr is reported to have replied, after the rise of quantum physics but before QM was discovered, to H. Høffding's question "Where can the photon be said to be?" with "To be, to be, what does it mean to

be?" (cited in [Wheeler and Ford 1998, p. 131]). Bohr might have been echoing the most famous sentence of Shakespeare's *Hamlet*, "To be, or not to be, that is the question" (Act 3, ll. 1749), realizing that in quantum physics one might want to or even must ask first "What does it mean to be?" (Hamlet's famous monologue is, too, about much more than merely deciding to live or die.) Høffding's and Bohr's questions are still unanswered and, in Bohr's ultimate, RWR-type, view, are unanswerable, when it comes to quantum objects, such as photons. Even as invisible to thought, quantum objects are idealizations (and hence still products of thought), in the present interpretation, ultimately only applicable at the time of observation, even if Bohr himself did not go that far. Either way, such questions as "Where can something be said to be?" or "When had something happened?" can only be asked about quantum phenomena, observed in measuring instruments, and as such visible to thought, to our mind's eye, or even to our immediate perceptuon. Nature has no photons or electrons, any more than being or reality, including that of the RWR-type. Admittedly, as are our thought and hence these concepts are created (we don't know how either) by our brains, which are part of by biological and neurological constitution, and in this sense still by nature. This is, however, not the same as saying (as is done sometimes) that nature uses these concepts through us. Rather, nature allows us to create concepts—daily, physical, philosophical, or mathematical—and use them in considering our interactions with nature by means of technology, beginning with that of our bodies, and again, our thought. It is this interaction and only this interaction that enables us to idealize some part of the constitution of nature, even its ultimate constitution, as something that is invisible to thought and hence that cannot appear in our mind's eye.

This brings me to my second epigraph "To die for the invisible—this is metaphysics," courtesy of E. Levinas's book, *Totality and infinity: An essay on exteriority* [Levinas 2012, p. 39] (originally published in French in 1961). The book is about ethics (as is, along with its many other themes, *Hamlet* as well), and as such it might appear distant from quantum physics. Levinas's epistemology, however, advanced in this book not only shares some of the philosophical genealogy, for example, in Kant's philosophy, with quantum theory, but might have more direct connections with it. Quantum theory and its epistemological problems, and possibly Bohr's ideas, were known to Levinas, as they were widely discussed on the French intellectual scene to which Levinas's work belongs. Levinas's concept of exteriority, expressly associated by him with "the invisible" [*l'invisible*], has manifested affinities with the idea of invisible to thought and even strictly means, in the ethical domain, that which is invisible to thought. My main interest here is the association, quite dramatic—"To die for the invisible!" No less!—of the invisible with metaphysics, which I would like to connects to quantum physics and indeed to all modern physics, from Galileo on, as a mathematical-experimental science. Insofar as one means by metaphysics something exterior to nature or, to use the ancient Greek word, *physis*, especially referring by metaphysics to something theological, modern physics excludes metaphysics. There is *no* metaphysics. On the other hand, insofar, because the idea of physics as a mathematical-experimental science or the idea of nature in the first place still belongs to thought, even when nature is invisible to thought, there is *only* metaphysics. Modern physics navigates and negotiates between both, "no metaphysics" and "only metaphysics." How close we come, in modern physics, to understanding nature, including in its ultimate constitution, even if the latter is ultimately invisible to thought, depends on our interactions with nature by means of experimental technologies and mathematics. These interactions are part of nature, too, but a particular part of it, specific to us, to our thinking and technologies, beginning with that of our bodies and brains, which are responsible for our thought. Our thought, however, also has a capacity to reach what is beyond it, is invisible to it, and to affirm it: to die for the invisible—this is metaphysics.

**Acknowledgments.** This paper was in part prompted by the question by Lorenzo Maccone concerning the relationships between the cat paradox and Bohr's complementarity, and the subsequent exchanges, for which I thank him, even though our views concerning quantum theory are very different. I am grateful to G. Mauro D'Ariano for exceptionally illuminating conversations concerning the subjects considered the article and fundamental physics in general. I am also happy to thank Gregg Jaeger and Andrei Khrennikov for many discussions on quantum foundations.